\documentclass[runningheads]{llncs}
\usepackage[T1]{fontenc}
\usepackage{graphicx}
\usepackage{booktabs}
\usepackage[misc]{ifsym}

\usepackage{hyperref}
\usepackage{cleveref}
\usepackage{booktabs}
\usepackage{multirow}
\usepackage{tablefootnote}
\usepackage{array}
\usepackage{amssymb}
\usepackage{caption}
\usepackage{subcaption}
\usepackage{placeins}
\usepackage{color,soul}
\usepackage{placeins} 

\newcommand{\para}[1]{\smallskip\noindent\textbf{#1}}

\begin{document}

\title{A Comparative Analysis of Wealth Index Predictions in Africa between three Multi-Source Inference Models\thanks{\color{red}Please cite the ECMLPKDD'24 version of this paper}}

\titlerunning{A Comparative Analysis of Wealth Index Predictions}

\author{Márton Karsai\inst{1,3} \orcidID{0000-0001-5382-8950} \and
János Kertész\inst{1,2} \orcidID{0000-0003-4957-5406} \and
Lisette Espín-Noboa\inst{2,1} (\Letter) \orcidID{0000-0002-3945-2966}}

\authorrunning{Karsai et al.}

\institute{Central European University, Vienna 1100, Austria \email{\{KarsaiM,KerteszJ,EspinL\}@ceu.edu}
\and
Complexity Science Hub, Vienna 1080, Austria
\and
National Laboratory for Health Security, HUN-REN Alfréd Rényi Institute of Mathematics, Budapest 1053, Hungary}
\maketitle              %

\begin{abstract}
Poverty map inference has become a critical focus of research, utilizing both traditional and modern techniques, ranging from regression models to convolutional neural networks applied to tabular data, satellite imagery, and networks. While much attention has been given to validating models during the training phase, the final predictions have received less scrutiny. In this study, we analyze the International Wealth Index (IWI) predicted by Lee and Braithwaite (2022) and Espín-Noboa et al. (2023), alongside the Relative Wealth Index (RWI) inferred by Chi et al. (2022), across six Sub-Saharan African countries. Our analysis reveals trends and discrepancies in wealth predictions between these models. 
In particular, significant and unexpected discrepancies between the predictions of Lee and Braithwaite and Espín-Noboa et al., even after accounting for differences in training data. In contrast, the shape of the wealth distributions predicted by Espín-Noboa et al. and Chi et al. are more closely aligned, suggesting similar levels of skewness.
These findings raise concerns about the validity of certain models and emphasize the importance of rigorous audits for wealth prediction algorithms used in policy-making. Continuous validation and refinement are essential to ensure the reliability of these models, particularly when they inform poverty alleviation strategies.

\end{abstract}
\section{Introduction}
Accurate and detailed poverty maps are indispensable for targeting anti-poverty interventions. While the United Nations prioritizes poverty eradication~\cite{UN_SDG1}, official poverty maps often do not exist~\cite{serajuddin2015data} or lack the granularity to identify specific areas of critical need due to data aggregation at broad administrative levels~\cite{ubos2019poverty}.
In recent years, predicting poverty using novel data sources has garnered significant attention within the AI for social good community. Diverse approaches leveraging satellite imagery~\cite{ghosh2010shedding,jean2016combining}, mobile phone data~\cite{cruz2021estimating,dong2014inferring}, social media activity~\cite{levy2019optimal,piaggesi2022mapping}, or combinations thereof~\cite{chi2022microestimates,lee2022high,espin2023interpreting}, have emerged, raising questions about the optimal methodology for accurate poverty estimation. Towards this goal, Espín-Noboa et al. 2023~\cite{espin2023interpreting} examined the performance of three machine learning architectures utilizing satellite imagery, tabular data, and a hybrid approach. Their findings revealed a nuanced landscape, with image-based models excelling at inferring the wealthiest areas, while tabular features proved more effective in rural regions. These findings highlight the importance of tailoring models to specific regions and feature sources.

However, current methods for creating poverty maps have some limitations. 
Although model architectures, features, and train-test validations are meticulously designed and evaluated on small samples, they may not ensure accurate predictions in the final poverty maps.
The lack of ground-truth data prevents the validation of these predictions, leading to potential inaccuracies. 
Consequently, these unvalidated predictions are used for policy-making, relying on training samples that may not accurately represent the entire population.

Another critical issue is the choice of ground-truth indicators.
Various studies use socioeconomic indicators such as income per capita~\cite{ebener2005wealth}, housing prices~\cite{wang2021deep}, and wealth indexes~\cite{chi2022microestimates,lee2022high,espin2023interpreting}, each with its strengths and limitations~\cite{poirier2020approaches}. 
Different wealth indexes such as the Relative Wealth Index (RWI)~\cite{rutstein2004dhs}, the Comparative Wealth Index (CWI)~\cite{rutstein2014making}, the Harmonized Wealth Index (HWI)~\cite{staveteig2014intertemporal}, and the International Wealth Index (IWI)~\cite{smits2015international}, provide distinct perspectives on wealth but also come with inherent constraints~\cite{lee2022high}. 
This variability makes it challenging to determine the most reliable inferred poverty map when comparing methodologies with similar predictive power.

If the validations of these different models are equally accurate, their resulting poverty maps should theoretically be similar. 
However, any discrepancies would prompt an investigation into the reasons for these differences and an assessment of which model is more accurate.
In this study, we aim to address this gap by comparing three recent wealth inference approaches: \textbf{IWI} by Lee and Braithwaite~\cite{lee2022high}, and by Espín-Noboa et al.~\cite{espin2023interpreting}, and \textbf{RWI} by Chi et al.~\cite{chi2022microestimates}.
We show their inferred poverty maps across six African countries and examine the degree of concordance %
of these predictions with established GDP trends, aiming to evaluate the robustness and reliability of these methodologies.
The analysis is two-fold. 
First, we compare the predicted distributions of wealth to assess the overall poverty inferred for each country by each model.
Second, considering the differences in geo-located places each approach uses in the final poverty map, we verify whether overlapping locations across methods receive consistent predictions and align with expected poverty trends.
In adherence to open science practices, we make both the code and the predictions openly available~\cite{espindata}.

\section{Preliminaries}
In this study, we compare the predictions of high-resolution poverty maps across six African countries from three different studies: \textbf{M1}~by Lee and Braithwaite (2022)~\cite{lee2022high}, \textbf{M2}~by Espín-Noboa et al. (2023)~\cite{espin2023interpreting}, and \textbf{M3}~by Chi et al. (2022)~\cite{chi2022microestimates}.

\subsection {Ground-Truth Data}
\label{sec:gt}

Ground-truth data in {M1}, {M2}, and {M3}~are derived from nationally-representative, population-based household surveys conducted by the Demographic and Health Survey (DHS) Program~\cite{dhs}. 
These surveys encompass asset/wealth information and sub-regional geomarkers. 
Each household survey includes numerous questions on socioeconomic status.
From the responses, DHS calculates the Relative Wealth Index (RWI)~\cite{cordova2009methodological,dhs_wealth_index} using the first principal component derived from a Principal Component Analysis (PCA)~\cite{hotelling1933analysis,jolliffe2016principal} of a standardized set of 15 questions concerning assets and housing characteristics. These questions cover various aspects such as electricity, telephone, automobile, refrigerator, TV, water supply, toilet facilities, and construction materials. 
This RWI serves as the ground-truth measure of wealth in {M3}.
In contrast, {M1}~and {M2}~compute the International Wealth Index (IWI)~\cite{smits2015international}, a similar metric, which assesses household material well-being and economic status across low and middle-income countries. 
The IWI is derived from 12 housing characteristics; {M1}~excludes two of them, while {M2}~uses them all. 
PCA is used to determine asset weights, which are re-scaled to produce IWI scores ranging from 0 to 100, representing the spectrum from no assets to the highest wealth. 
The IWI's advantage lies in its comparability across different countries and years, unlike the country-specific RWI used in {M3}.

The studies also differ in their spatial resolution. 
{M1}~and {M2}~use actual DHS cluster locations which typically contain 25-30 households, and are geo-located with added noise to protect respondents privacy. The noise varies: up to 2 km for urban and 5 km for rural clusters, with 1\% of rural clusters displaced by up to 10 km~\cite{burgert2013geographic}.
Conversely, {M3}~employs a 2.4 km grid system defined by Bing tile coordinates to map DHS clusters to grid cells.

\begin{table}[t]
    \centering
    \caption{\textbf{Years of collected DHS ground-truth and features per study.} 
    Instances where two or more models used the same ground-truth data are few: {M1}~and {M2}~in Uganda, {M1}~and {M3}~in Rwanda, and all models in South Africa. 
    {M1}~inferred Gabon's poverty map using a model trained on data from 25 Sub-Saharan African countries, as the most recent survey from Gabon in 2012 was considered outdated.
    The table also categorizes the features utilized by each model. 
    Note that for this analysis, unlike the original models reported in~\cite{espin2023interpreting}, the {M2}~models were retrained from scratch without mobility and demographic features, as the mobility data was discontinued and the Facebook Marketing API redefined its reach estimates. This change did not affect performance.
    }
    \label{tbl:ground_truth}
    \begin{tabular}{@{}p{5.8cm}%
    >{\raggedleft\arraybackslash}p{2.cm}%
    >{\raggedleft\arraybackslash}p{2.cm}%
    >{\raggedleft\arraybackslash}p{2.cm}@{}}
    \toprule
    Country (ISO 3166-1 alpha-3 Code)      & {M1}        &        {M2}  &         {M3}  \\ \midrule
    Sierra Leone (SLE) &          2016 &     2016,2019 &     2013 \\
    Liberia (LBR)      &          2016 &     2019,2022 &     2013 \\
    Uganda (UGA)       &  2016,2018-19 &  2016,2018-19 &     2016 \\
    Rwanda (RWA)       &       2014-15 &          2019 &  2014-15 \\
    South Africa (ZAF) &          2016 &          2016 &     2016 \\
    Gabon (GAB)        &          Multiple &          2019 &     2012 \\
    \toprule
    Feature                              &         {M1}        &        {M2}*  &         {M3}  \\ \midrule
    Daylight satellite images            & \checkmark & \checkmark & \checkmark \\
    Nightlight intensity                 & \checkmark & \checkmark & \checkmark \\
    Infrastructure                       & \checkmark & \checkmark & \checkmark \\
    Connectivity                         &          - & \checkmark & \checkmark  \\
    Population                           & \checkmark & \checkmark & \checkmark  \\
    Mobility network $\dagger$           &          - & (\checkmark) &          -  \\
    Demographics of people $\dagger$     &          - & (\checkmark) &          -  \\
    Natural conditions                   &          - &          - & \checkmark  \\
    \midrule
    Number of total features $\ddagger$             &       249+4 &   173+784 & 12+100    \\
    \bottomrule
    \end{tabular}
\\
\begin{tabular}{@{}l@{}}
\scriptsize{* This model has three variants: CNN (only images), CatBoost (all but images), and combined (all features).} \\
\scriptsize{$\dagger$ We exclude these features due to discontinued support (mobility) and rapid changes of API (demographics).} \\
\scriptsize{$\ddagger$ Metadata (non-imagery) features (249, 173, and 12 resp.) plus image-based features (4, 784, and 100 resp.).} \\
\end{tabular}
\end{table}

The final poverty map is defined by the populated places (i.e., cities, towns, villages, hamlets and isolated dwellings) extracted from OpenStreetMap (OSM) in {M1}~and {M2}~({M2}~additionally includes neighborhoods), and by the grid-cells covering 
the entire country %
in {M3}.

Another critical difference among the three studies is the year in which the surveys were conducted. Each study utilized training data from different years (see~\Cref{tbl:ground_truth}), complicating the direct comparison of their results. Consequently, as explained in~\Cref{sec:trends}, our comparison will not concentrate on the precise predictions of each study. Instead, we will focus on identifying expected trends, using the Gini coefficient of wealth distributions and GDP per capita as baselines. These baselines are sourced from official data provided by the World Bank~\cite{world_bank_gdp_per_capita,world_bank_gini}. This approach allows us to account for temporal variations and provides a more meaningful analysis of the wealth and economic status trends across different periods.

\begin{figure}[t!]
    \centering
    \includegraphics[width=0.98\textwidth]{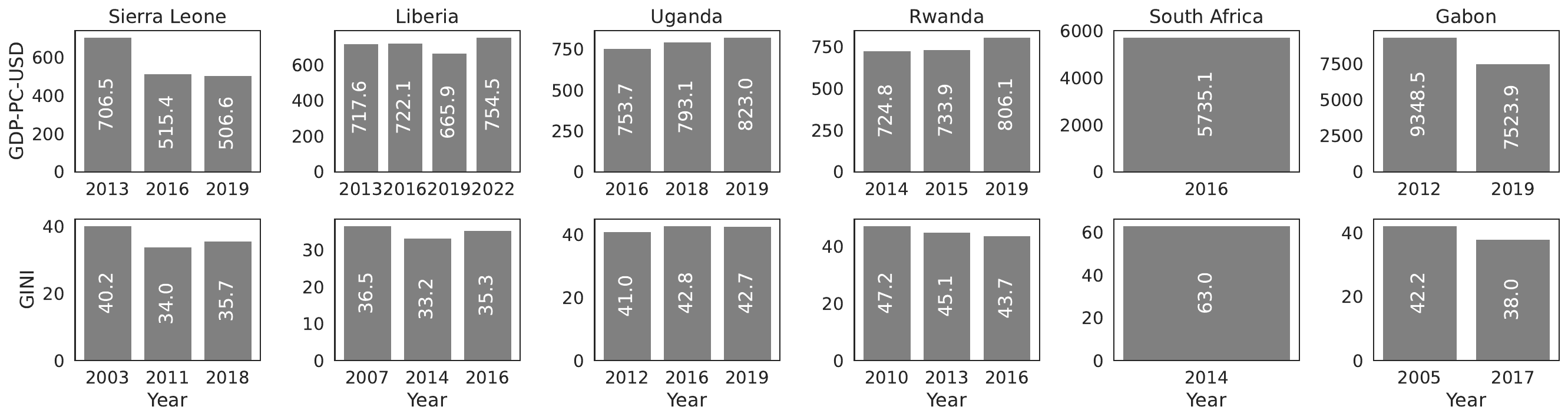} %
    \vspace{-2ex}
    \caption{\textbf{Available GDP per capita and Gini index.}
    Top: GDP values are aligned with the years of the surveys used in the models (x-axis).
    Bottom: The most recent available Gini indices are shown for reference.
    Uganda is the only country with GDP and Gini index data available for approximately the same years as the surveys.
    Both metric values are provided by the World Bank~\cite{world_bank_gdp_per_capita,world_bank_gini}.
    }
    \label{fig:preliminaries}
\end{figure}

\subsection{Models}
\label{sec:models}

\para{{M1}:}
IWI inference model by Lee and Braithwaite 2022~\cite{lee2022high} combines feature-based and image-based models to enhance each other's performance. It uses two cross-validation approaches: 
5-fold cross-validation (basic CV) and leave-one-country-out cross-validation (leave-country-out). 
Initially, an eXtreme Gradient Boosting (XGBoost) algorithm is trained on $249$ features (infrastructure, night-time luminosity, and population density) to predict IWI scores. Bayesian optimization is employed to tune the hyperparameters, selecting the best configuration for the training dataset. The XGBoost model's predictions are then used to train a Convolutional Neural Network (CNN) model that outputs a probability distribution of socioeconomic status (rich, upper-middle class, lower-middle class, poor) for a 1-square-mile area. These probabilities are then fed back into the XGBoost model, creating a cycle where the models iteratively improve by providing better training data.

\para{{M2}:}
IWI inference model by Espín-Noboa et al. 2023~\cite{espin2023interpreting} extracts $957$ features from seven data sources. Each location is defined by the centroid of each place, using a default bounding-box width of 1 mile. Some features are also queried within 2, 5, and 10 Km to capture the original locations of the DHS clusters, similar to {M1}. 
Features include high-resolution satellite imagery, demographics, mobility networks, and mobile phone antennas.
Three machine learning models are proposed to predict both the mean and standard deviation of IWI scores, providing insights into the wealth distribution within clusters. The first model is a CNN that predicts IWI scores from daylight satellite images. This CNN comprises 22 layers, with the final layer using a linear activation function and mean-squared error (MSE) as the loss function, tuning 4 hyperparameters. The second model is a CatBoost regressor (CB), predicting IWI scores from 173 metadata features and evaluated individually for each data source. This model is tuned using 11 hyperparameters. The third model combines CNN+CB by feeding the third-to-last layer of the CNN into the CB model, adding 784 features. For training, data is partitioned into 80\% train and 20\% test sets, stratified into 10 equal-width bins of wealth. A 4-fold cross-validation on the train set, stratified by SES (ses CV), is used to tune hyperparameters via Random Search on 200 combinations. This process is repeated three times independently with different random seeds, to control for fluctuations, and mean performance is reported. The models are evaluated on various configurations of training data, including different years of ground-truth data, relocation strategies, data augmentation ($*_a$), sample weight strategies ($*_w$), and feature sources, ensuring robust and comprehensive performance evaluation.
For our analysis, we chose the model with the best performance (lowest root-mean-square error, RMSE) in each country:
CB for Sierra Leone and Liberia, CB$_w$ for Uganda and Rwanda, and CNN$_a$+CB for Gabon and South Africa. %

\para{{M3}:} RWI inference model by Chi et al. 2022~\cite{chi2022microestimates,facebook_relative_wealth_index} utilizes $112$ features in total, including high-resolution satellite imagery, mobile phone network data, topographic maps, and aggregated, de-identified connectivity data from Facebook. These diverse data sources are processed through deep learning and computational algorithms to generate quantitative features for each village. These features are then used to train a gradient-boosted regression tree, a supervised machine-learning model, to predict RWI scores. Hyperparameters for the gradient-boosted tree are tuned to minimize the cross-validated MSE, using a grid search. They use three cross-validation approaches, two of which are also employed by {M1}: 5-fold cross-validation (basic CV), leave-one-country-out cross-validation (leave-country-out), and spatially stratified cross-validation (spatial CV). The use of different cross-validation methods helped highlight the potential upward bias in performance due to spatial auto-correlation in the training and test data. All analysis in this study is based on spatial CV, ensuring conservative and appropriate model training for geographic data with spatial auto-correlation.

In summary, all models use a multi-modal and hybrid approach, combining features from various data sources and two model architectures, see~\Cref{tbl:ground_truth} (bottom). {M2}~and {M3}~use satellite images by pre-training a CNN model and feeding the image embeddings into a gradient boosting algorithm. {M1}~also uses a hybrid approach but first pre-trains a gradient boosting algorithm to train a CNN, and then creates a feedback loop that enhances the initial model. Only {M3}~estimates the expected error of each location, though its explanatory power is limited (average $R^2 = 0.09$). All models predict the mean wealth of locations, but only {M2}~additionally predicts their standard deviation, and proposed two additional models to separate the effects of images and metadata.

\subsection{Expected Trends} %
\label{sec:trends}
Before digging into the comparison of predictions between the three models of interest~\cite{chi2022microestimates,lee2022high,espin2023interpreting}, it is crucial to acknowledge the discrepancies in the ground-truth data used to train them. 
\Cref{tbl:ground_truth} (top) outlines the specific years for which each approach utilized data for each country. 
This variation in time is significant because wealth and poverty are dynamic concepts that can fluctuate over time. 
To address this, rather than comparing directly the predicted values, we focus on the trends in predictions. 
We will achieve this by examining the trends of changes in Gross Domestic Product (GDP) per capita provided by the World Bank~\cite{world_bank_gdp_per_capita},
as depicted in~\Cref{fig:preliminaries} (top). 
By analyzing these trends, we can hypothesize that the comparison between the models will mirror the trends observed in GDP. %
This approach allows us to account for potential discrepancies arising from  the distinct time-frames of the datasets. %

Additionally, we compare the changes in the heterogeneity of the predicted wealth distributions with each country's official Gini index. Due to the lack of data from the World Bank~\cite{world_bank_gini},
for the same years as the surveys in most countries, see~\Cref{fig:preliminaries} (bottom), we present comparisons only for Uganda, where data is available, and South Africa, where all models used the same ground-truth. %

For our analysis, we perform pairwise comparisons between models, independently comparing {M3}~with both {M1}~and {M2}. When comparing {M3}~with {M1}, we apply a linear transformation to the RWI scores, converting them to their IWI equivalents in {M1}~due to their strong correlation (Pearson's correlation coefficients: $0.93-0.99$~\cite{lee2022high}). Similarly, we perform the corresponding transformation using {M2}~as the baseline. It is important to note that these transformations are approximate, as the RWI is inherently a relative score. Nevertheless, given the high-correlation between RWI and IWI, this assumption is justified. Future research could explore re-training {M3}~to directly predict IWI scores or transforming RWI into CWI, although this method remains imprecise~\cite{rutstein2014making}. The results comparing {M3}~with {M1}~and {M2}, presented in~\Cref{app:chi},
should therefore be interpreted with caution.

\begin{table}[t]
\centering
\caption{\textbf{Prediction statistics.} We show the number $N$ of target locations contained in the inferred poverty maps, and the mean, standard deviation (SD),  Gini coefficient, and skewness test (SK)~\cite{2020SciPy-NMeth} of the predicted wealth scores for each model. 
None of the distributions are normal (SK test, $p < 0.001$). Gini for {M3}~is omitted due to negative RWI values.
}
\label{tbl:summary_predictions}
\begin{tabular}{@{}llrrrrrlrrrrrlrrrr@{}}
\toprule
\multirow{2}{*}{Country's} && \multicolumn{5}{c}{{M1}~(IWI)} && \multicolumn{5}{c}{{M2}~(IWI)} && \multicolumn{4}{c}{{M3}~(RWI)} \\ 
\cmidrule(lr){3-7} \cmidrule(l){9-13} \cmidrule(l){15-18} 
 Code
 && $N$ & Mean & SD & Gini & SK
 && $N$ & Mean & SD & Gini & SK
 && $N$ & Mean & SD & SK \\ 
 \midrule
SLE   &       &  13040 &    18.93 &   10.34 &    23.01 &             72.15 &       &   9881 &    21.87 &    6.09 &    11.54 &             70.52 &       &   8435 &    -0.35 &    0.29 &             44.13 \\
LBR   &       &  16525 &    14.57 &    8.79 &    25.93 &             90.74 &       &  15597 &    23.35 &    3.74 &     7.75 &             75.15 &       &   6189 &    -0.38 &    0.33 &             36.12 \\
UGA   &       &  27482 &    22.27 &   11.26 &    25.09 &             87.23 &       &  27791 &    41.83 &    5.21 &     5.95 &             86.16 &       &  25404 &    -0.23 &    0.31 &             65.00 \\
RWA   &       &  23323 &    24.40 &   14.51 &    28.96 &             75.31 &       &   1150 &    47.79 &    3.46 &     3.74 &             13.65 &       &   3715 &    -0.04 &    0.36 &             24.85 \\
ZAF   &       &  22687 &    61.14 &   18.10 &    16.85 &              4.12 &       &   3043 &    52.65 &   14.11 &    15.04 &             12.40 &       &  39072 &    -0.06 &    0.48 &             57.46 \\
GAB   &       &   4835 &    21.89 &    8.86 &    17.93 &             46.60 &       &   1135 &    48.32 &    6.72 &     7.38 &             12.61 &       &   1110 &    -0.24 &    0.44 &             12.13 \\
\bottomrule
\end{tabular}
\end{table}

\begin{table}[t]
\centering
\caption{\textbf{Expected vs. observed wealth differences.} %
Significant differences in predicted wealth between models are observed in LBR, UGA, RWA, and GAB, with {M2}~consistently predicting wealthier regions. RWA followed the expected trend but was notably over-represented. 
Identical wealth and Gini values were expected for UGA and ZAF due to the shared years of ground-truth data. However, {M1}~under-estimated wealth in UGA and slightly overestimated it in ZAF. In both cases, M1 generated more heterogeneous distributions compared to {M2}.
All differences are statistically significant ($p < 0.001$, Kolmogorov-Smirnov test~\cite{hodges1958significance}). \textbf{Highlighted values} show the correct trend (+/-). No available data (NA) for GAB (trained on data from other countries/years~\cite{lee2022high}) and for SLE, LBR, and RWA (due to the absence of official data). 
}
\label{tbl:summary_differences}
\begin{tabular}{@{}llrrrrrr@{}}
\toprule
\multirow{2}{*}{Country's Code} && \multicolumn{5}{c}{{M1}~vs. {M2}} \\ \cmidrule(l){3-7} 
 && \begin{tabular}[c]{@{}r@{}}Exp.\\IWI\end{tabular} & \begin{tabular}[c]{@{}r@{}}Obs.\\IWI\end{tabular} & \begin{tabular}[c]{@{}r@{}}Exp.\\Gini\end{tabular} & \begin{tabular}[c]{@{}r@{}}Obs.\\Gini\end{tabular} & KS \\ \midrule
SLE   &       &                        0.00 &                       -0.07 &                     NA &                     11.47 &  0.62 *** \\
LBR   &       &                        0.01 &                       -0.23 &                     NA &                     18.19 &  0.83 *** \\
UGA   &       &                        0.00 &                       -0.31 &                        0 &                     19.15 &  0.87 *** \\
RWA   &       &                       \textbf{-0.05} &                       \textbf{-0.32} &                     NA &                     25.22 &  0.85 *** \\
ZAF   &       &                        0.00 &                        0.07 &                        0 &                      1.81 &  0.30 *** \\
GAB   &       &                         NA &                       -0.38 &                     NA &                     10.55 &  0.95 *** \\
\bottomrule
\end{tabular}
\\
\begin{tabular}{@{}p{12cm}@{}}
\scriptsize{Expected IWI Difference $ = (GDP_{Y_{M1}} - GDP_{Y_{M2}})/(GDP_{Y_{M1}} + GDP_{Y_{M2}})$} \\
\scriptsize{Observed IWI Difference $ = (\mu_{IWI_{M1}} - \mu_{IWI_{M2}})/(\mu_{IWI_{M1}} + \mu_{IWI_{M2}})$}\\
\scriptsize{Expected Gini Difference $ = (Gini_{Y_{M1}} - Gini_{Y_{M2}})$} \\
\scriptsize{Observed Gini Difference $ = (Gini_{M1} - Gini_{M2})$}\\
\scriptsize{%
$Y_{M_*}$ denotes the year of the ground-truth data used to train model $M_*$, which is then used to select the corresponding indicator, GDP per capita or Gini coefficient, from the World Bank.}
\end{tabular}
\end{table}

\section{Results}

Here, we compare the inferred poverty maps produced by {M1}, {M2}, and {M3}.
We first present the predicted overall wealth distributions. %
Due to the lack of ground truth in all places of these poverty maps, we compare the predicted distributions with official GDP and Gini coefficient trends to understand how well these predictions align with established economic indicators. 
Given that each study employs different spatial resolutions, our comparison focuses on locations that overlap within each pair of models (e.g., {M1}~vs. {M2}).
Note that these analyses are based on the predicted poverty maps rather than test sets with ground truth. 
We trained all variants of {M2}~models and use the best model in each country based on their lowest RMSE scores (see details in~\Cref{sec:models}) to infer the final poverty maps.
The inferences of all models are publicly available: {M1}~\cite{DVN/5OGWYM_2022}, {M2}~\cite{espindata}, and {M3}~\cite{chidata}.
For a comparison of the reported predictive power in terms of R$^2$ for these studies on their respective test sets, please refer to~\Cref{app:performance}.

\subsection{Predicted wealth distribution}
\Cref{tbl:summary_predictions} summarizes the poverty map inferences for each country and model. First, we observe that {M1}~generally identifies more locations (OSM populated places) compared to {M2}~and {M3}. 
Second, all models produce skewed distributions (indicating high wealth inequality), as reflected in their high Gini coefficients and significant skewness test. However, when comparing {M1}~with {M2}, we find that {M1}~generates more skewed distributions (higher Gini) and predicts lower mean wealth in most cases, except for South Africa.

\begin{figure}[t]
     \centering
     \begin{subfigure}[b]{0.6\textwidth}
         \centering
         \includegraphics[width=\textwidth]{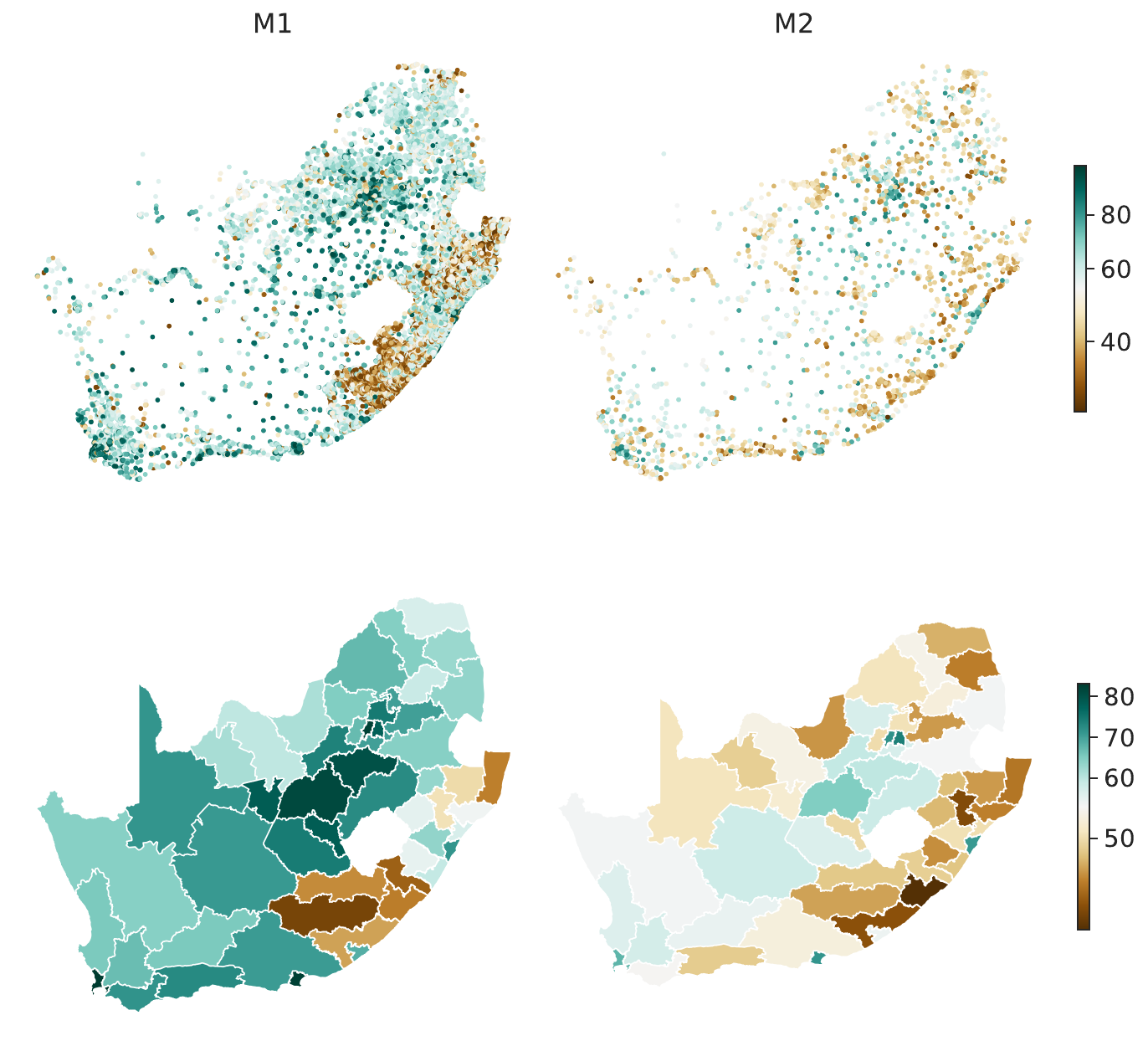}
         \caption{Poverty Map}
         \label{fig:ZAF_hr}
     \end{subfigure}
     \begin{subfigure}[b]{0.39\textwidth}
         \centering
         \includegraphics[width=\textwidth]{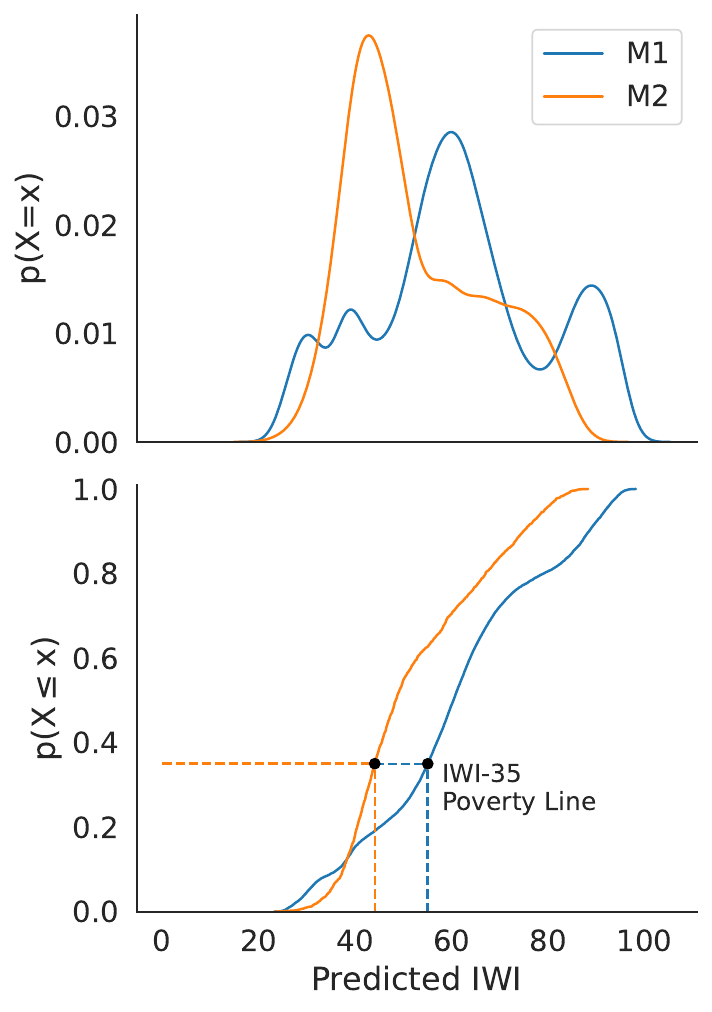}
         \caption{Wealth distribution}
         \label{fig:ZAF_admin}
     \end{subfigure}
        \caption{\textbf{IWI predictions in South Africa, ZAF ({M2}: CNN$_a$+CB).} 
        Both models were trained on the same ground-truth data, thus, we expect similar results.
        (a) Poverty maps.
        Color bars display the average predicted IWI scores, centered at the mean of {M2}'s predictions for comparison.
        Top: Prediction at the original resolution, OSM populated places. Both {M1}~and {M2}~include the Marion Island, omitted here for visualization purposes.
        Bottom: Aggregated mean wealth at the district level (administrative level 2).
        Compared to {M2}, {M1}~predicts higher wealth.
        (b) Predicted wealth distribution (top: PDF, bottom: CDF). 
        ZAF is the second country with the highest GDP in our study (see~\Cref{fig:preliminaries}), and this is reflected in the prediction by both models (e.g., $\mu>48$).
        {M2}~predicts a skewed, unimodal distribution of wealth, while {M1}~predicts an unusual multimodal distribution.
        IWI poverty line ($35^{th}$ percentile, corresponding to poverty headcount ratios at \$1.25 a day~\cite{smits2015international}) slightly differs between models.
        }
        \label{fig:ZAF}
\end{figure}

\subsection{Expected vs. Obtained Trends}
We compared {M1}~to {M2}, focusing on their expected performance based on changes in GDP per capita for each  country over the years of each model's data.\footnote{We refer the reader to~\Cref{app:chi} for approximate comparisons with {M3}.}  %
Results are shown in~\Cref{tbl:summary_differences}.
In Uganda, where both models used the same ground-truth, we expected similar wealth predictions. However, {M1}~inferred lower and more heterogeneous wealth compared to {M2}.
A similar expectation applied to South Africa, but while the results were comparable on average, {M1}~predicted wealthier places and more heterogeneous. %
In Liberia, {M1}~used $2016$ ground-truth data, while {M2}~used data from $2019$ and $2022$ (\Cref{tbl:ground_truth}).
During this period, GDP per capita dropped from $722.1$ in $2016$ to $665.9$ in $2019$, then rose to $754.5$ in $2022$ (\Cref{fig:preliminaries}). 
Averaging GDP between $2019$ and $2022$ (years used to train {M2}), we expected a slight decrease from $722.1$ to $710.2$. This is reflected in the normalized GDP difference ($0.01$), suggesting that {M2}'s IWI scores should be slightly lower than {M1}'s. However, {M2}~unexpectedly predicted higher values ($-0.23$).
In terms of Gini coefficients, {M1}~generally predicted more heterogeneous wealth distributions than {M2}.
Due to data limitations, we were unable to track Gini coefficient trends for the other countries.
These patterns are visualized in~\Cref{fig:ZAF} for South Africa, and \Cref{appendix_overall} for other countries.

\begin{table}[t]
\centering
\caption{\textbf{Prediction differences in overlapping places.}
Predictions are compared only for locations common to both models, defined as those within $500~m$ of each other. Duplicates were removed by retaining the closest pairs. As expected, {M2}~contains many locations from {M1}, as both use OSM populated places, though {M1}~includes more (\Cref{tbl:summary_predictions}). Predictions are most similar in SLE (low RMSE, high Pearson correlation), with larger differences observed in other countries. \textbf{Highlighted values} show the correct trend (+/-).
}
\label{tbl:summary_overlap}
\begin{tabular}{@{}llrrrrrrrrr@{}}
\toprule
\multirow{2}{*}{Country's Code} && \multicolumn{9}{c}{{M1}~vs. {M2}} \\ \cmidrule(l){3-11}  
 && \begin{tabular}[c]{@{}r@{}}$N_o$\end{tabular} & 
 \begin{tabular}[c]{@{}r@{}}$\%$ in\\{M1}\end{tabular} & 
 \begin{tabular}[c]{@{}r@{}}$\%$ in\\{M2}\end{tabular} & 
 \begin{tabular}[c]{@{}r@{}}RMSE\end{tabular} & \begin{tabular}[c]{@{}r@{}}Pearson\\Corr.\end{tabular} & \begin{tabular}[c]{@{}r@{}}Exp.\\IWI.\end{tabular} & 
 \begin{tabular}[c]{@{}r@{}}Obs.\\IWI.\end{tabular} & 
 \begin{tabular}[c]{@{}r@{}}Exp.\\Gini\end{tabular} &
 \begin{tabular}[c]{@{}r@{}}Obs.\\Gini\end{tabular}
 \\ \midrule
SLE   &       &             6227 &                 47.75 &                 63.02 &                6.63 &              0.80 *** &                                     0.00 &                                -0.10 &                             NA &                     8.89 \\
LBR   &       &            13786 &                 83.43 &                 88.39 &               11.00 &              0.68 *** &                                     0.01 &                                -0.27 &                             NA &                    14.46 \\
UGA   &       &            10981 &                 39.96 &                 39.51 &               19.96 &              0.78 *** &                                     0.00 &                                -0.28 &                                0 &                    19.31 \\
RWA   &       &              653 &                  2.80 &                 56.78 &               22.18 &              0.35 *** &                                    \textbf{-0.05} &                                \textbf{-0.22} &                             NA &                    21.04 \\
ZAF   &       &             2686 &                 11.84 &                 88.27 &               22.61 &              0.46 *** &                                     0.00 &                                 0.13 &                                0 &                     0.91 \\
GAB   &       &              581 &                 12.02 &                 51.19 &               25.13 &              0.73 *** &                                      NA &                                -0.32 &                             NA &                    18.21 \\     
\bottomrule
\end{tabular}
\end{table}

\begin{figure}[t]
     \centering
         \centering
         \includegraphics[width=\textwidth]{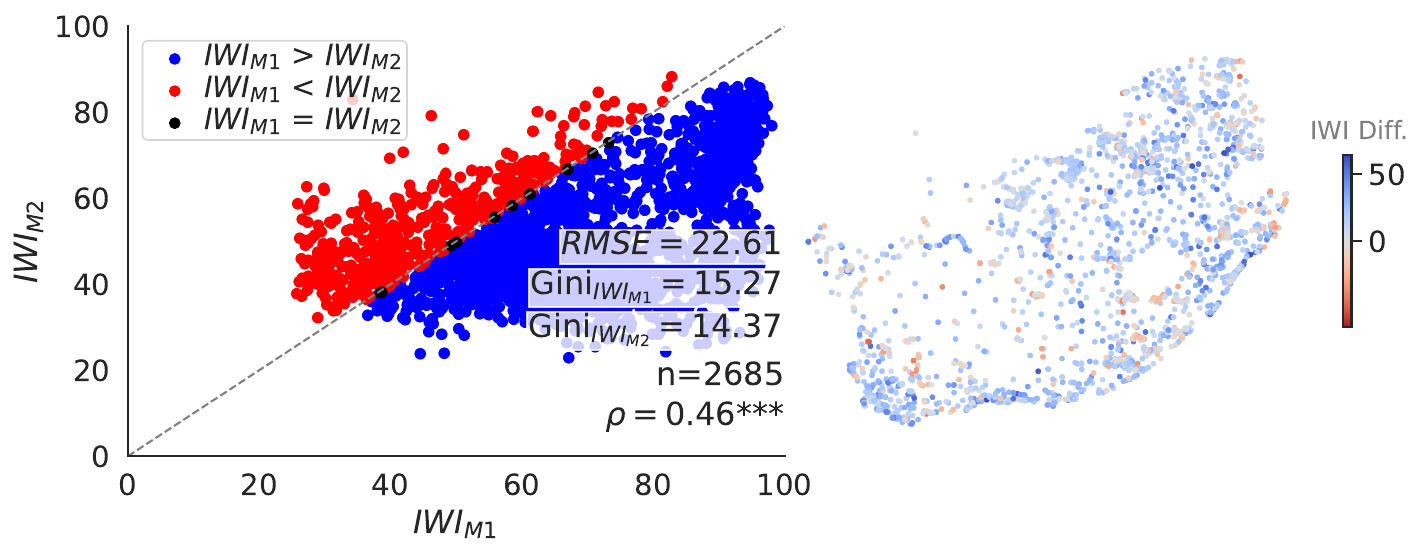}
        \caption{\textbf{IWI predictions at overlapping places in South Africa, ZAF ({M2}: CNN$_a$+CB).} 
        Right: IWI differences between the predictions of {M1}~and {M2}~for each overlapping place.
        Left: Correlation between these predictions.
        Colors indicate the direction of the differences: %
        {M1}~generally overestimates wealth compared to %
        {M2}~(mostly blue). Moreover, the wealth distribution in these locations is equally skewed (similar Gini coefficient). 
        }
        \label{fig:ol_ZAF}
\end{figure}

\subsection{Overlapping places}
The previous results compared overall predictions across models, offering a broad understanding of their performance relative to country-level GDP per capita and Gini coefficient trends. Now, we focus on comparing the predictions for locations that overlap in both {M1}~and {M2}. 
We perform a spatial join to analyze only these overlapping locations, which we expect to perfectly reflect the anticipated trends as both models rely on OSM populated places (see~\Cref{sec:gt}).

As before, we compare the expected GDP per capita differences with the observed variations in predicted mean wealth between the two models, see~\Cref{tbl:summary_overlap}. When comparing these findings with the overall predictions (\Cref{tbl:summary_differences}), we observe consistent trends in both wealth predictions and Gini coefficients.
In all countries, {M2}~predicts higher average wealth than {M1}, with the exception of ZAF. Additionally, {M1}~consistently produces more skewed distributions, indicating higher wealth inequality. However, the magnitude of these differences is generally smaller in overlapping locations compared to all locations (\Cref{tbl:summary_differences}). This suggests that while the trends in overlapping locations do not perfectly match expectations, the values are closer to what was anticipated.
For instance, in UGA and ZAF, where identical predictions were expected, the results align more closely with the anticipated outcome.
In RWA, as predicted, {M2}~identifies wealthier places, but with a slight over-estimation. The correlation between these predicted wealth scores for ZAF is shown in~\Cref{fig:ol_ZAF} 
(\Cref{appendix_overall} for other countries).

\section{Conclusions}

This study aimed to evaluate the alignment between predicted wealth trends from three models---{M1}~(Lee and Braithwaite, 2022~\cite{lee2022high}), {M2}~(Espín-Noboa et al., 2023~\cite{espin2023interpreting}), and {M3}~(Chi et al., 2022~\cite{chi2022microestimates})---using IWI and RWI scores as proxies for economic well-being. The comparison of these models centered on understanding their differences in underlying ground-truth data and their ability to predict wealth in relation to expected changes in GDP per capita and wealth inequality (Gini coefficient).

Our primary focus was on comparing IWI predictions from {M1}~and {M2}, as they directly provide comparable outputs. While {M3}~predicts RWI scores, the conversion to IWI for proper comparison remains complex and non-trivial. Although some prior studies~\cite{rutstein2014making,staveteig2014intertemporal} have proposed methods for these conversions, they come with inherent limitations. For the sake of simplicity and based on the high correlation between RWI and IWI scores (Pearson's correlation: $0.93-0.99$~\cite{smits2015international}), we included {M3}~in our comparison (see~\Cref{app:chi}), but the core analysis remained on {M1}~and {M2}.

Our findings indicate unexpected differences between the predictions of {M1}~and {M2}. {M2}~generally predicts wealthier regions, while {M1}~over-represents poverty. Based on GDP per capita trends during the training years, the models should produce similar wealth predictions, except in Rwanda, where the trend direction is captured but exaggerated. In Uganda and South Africa, where each model used the same years of ground-truth data, we expected similar wealth scores. However, {M1}~predicted lower wealth in Uganda and higher wealth in South Africa, contradicting our hypothesis of consistent distributions between the models.

In places where both models overlap, observed trends from the predictions also deviated from expected wealth trajectories (\Cref{tbl:summary_overlap}). However, these discrepancies were smaller compared to the ones observed in the overall predictions (\Cref{tbl:summary_differences}). In the overlapping places, {M1}~remained more heterogeneous than {M2}, suggesting greater inequality. The absence of official Gini coefficients for the relevant years prevented us from verifying whether the observed wealth inequality trends in Sierra Leone, Liberia, Rwanda, and Gabon reflect reality.

The observed discrepancies between model predictions highlight the critical need for further scrutiny before using these inferred poverty maps for policy-making. While model evaluations during training provide insight into test data performance, they do not guarantee the broader accuracy of the inferred wealth indicators. Accurate and reliable poverty maps are fundamental to informing effective policy interventions, particularly in line with the United Nations' first Sustainable Development Goal: eradicating global poverty.

Moving forward, more comprehensive audits of these models and their underlying assumptions are essential. Collaboration between AI practitioners, policymakers, and local stakeholders in African countries is crucial to ensure that poverty maps accurately reflect local realities. Future research should prioritize refining methodologies for converting RWI to IWI scores to enable more nuanced comparisons across wealth indices. Alternatively, a replication study using M3's architecture to predict IWI scores could offer more precise cross-model cross-country and cross-year comparisons. Governments, policymakers, and NGOs must prioritize providing up-to-date global statistics to enhance the evaluation of wealth inequality trends generated by machine learning and AI technologies, thereby supporting progress toward broader Sustainable Development Goals (SDGs).

\begin{credits}
\subsubsection{\ackname} This work was funded by the SoBigData++ project (H2020-871042), Dataredux ANR project (ANR-19-CE46-0008), and the CHIST-ERA project SAI: FWFI 5205-N. 
The presented computational results have been achieved in part using the Vienna Scientifc Cluster (VSC).
This work was partially funded by the Vienna Science and Technology Fund WWTF under project No. ICT20-079.
We thank the Data for Good team at Meta for insightful discussions during the first partner lightning talks, where early versions of this work were presented. 
We also appreciate the authors of the analyzed papers for their open data practices, which made this audit possible.

\subsubsection{\discintname}
The authors have no competing interests to declare that are
relevant to the content of this article.

\end{credits}

\bibliographystyle{splncs04}

\setcounter{table}{0}
\renewcommand{\thetable}{A\arabic{table}}

\setcounter{figure}{0}
\renewcommand{\thefigure}{A\arabic{figure}}

\appendix
\section{Appendix}
\label{appendix}
This appendix provides additional material comparing models {M1}~and {M2}~across five African countries: Sierra Leone, Liberia, Uganda, Rwanda, and Gabon. Due to the presence of negative RWI values in {M3}~and the complex conversion to IWI scores~\cite{rutstein2014making}, we compare the distribution shapes across all models and assess their similarity, along with the reported validation scores. This analysis complements our main findings by offering further insights into the models' performance and wealth estimates.

\para{Prediction data.} Openly available here: {M1}~\cite{DVN/5OGWYM_2022},  {M2}~\cite{espindata}, and {M3}~\cite{chidata}.

\subsection{{M1}~vs. {M2}~on remaining countries}
\label{appendix_overall}
In this section, we present the plots comparing the predictions of models {M1}~and {M2}~across the remaining five African countries, complementing the results presented in the main text.

\clearpage
\newpage
\subsubsection{Sierra Leone, SLE}
~
\vspace{-10pt}
\begin{figure}[h!]
     \centering
     \begin{subfigure}[b]{0.6\textwidth}
         \centering
         \includegraphics[width=\textwidth]{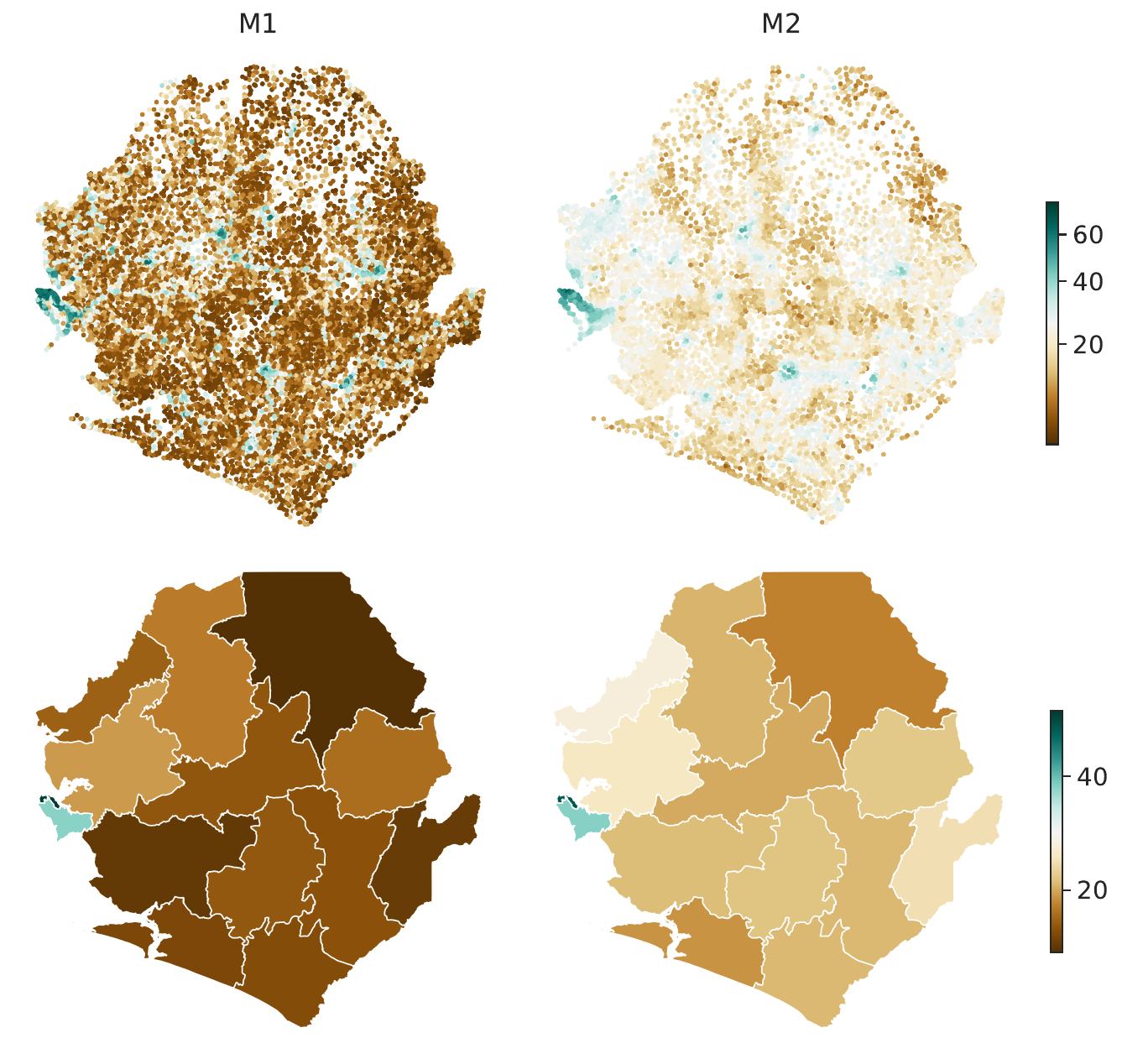}
         \caption{Poverty Map}
         \label{fig:SLE_hr}
     \end{subfigure}
     \begin{subfigure}[b]{0.39\textwidth}
         \centering
         \includegraphics[width=\textwidth]{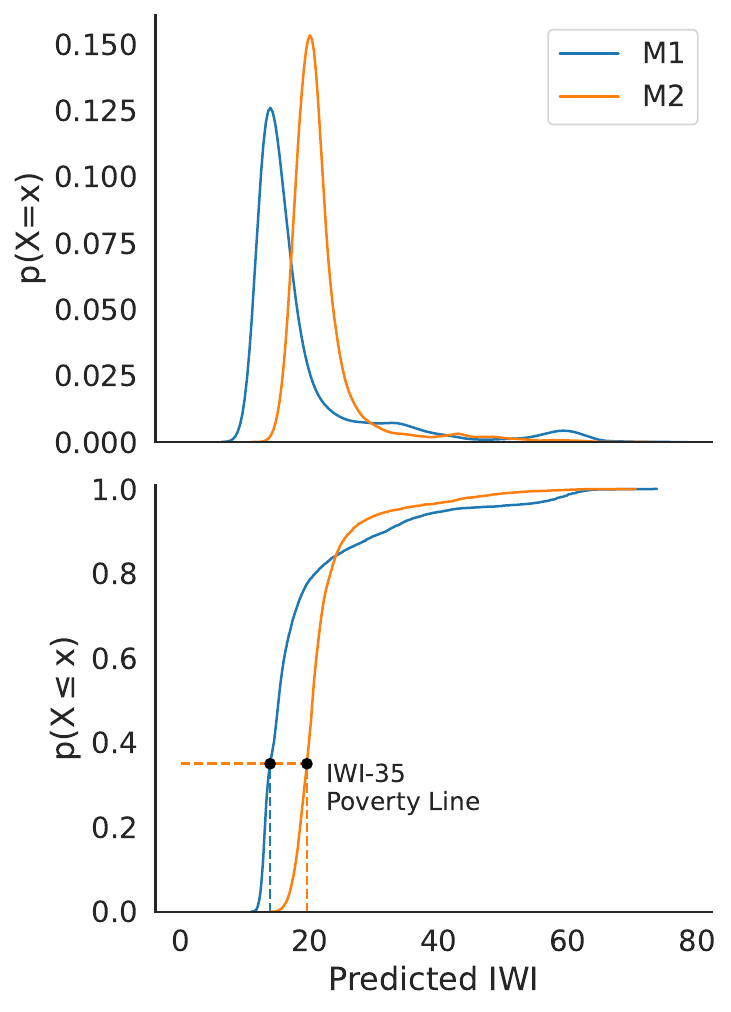}
         \caption{Wealth distribution}
         \label{fig:SLE_admin}
     \end{subfigure}
        \caption{\textbf{IWI predictions in SLE ({M2}: CB).} 
        (a) Poverty maps with colors representing the average predicted IWI scores. %
        Top: OSM populated places. %
        Bottom: District level. %
        {M1}~predicts lower wealth than {M2}.
        (b) Predicted wealth distribution. %
        IWI poverty line ($35^{th}$ percentile, corresponding to poverty headcount ratios at \$1.25 a day~\cite{smits2015international}) slightly differs between models.
        }
        \label{fig:SLE}
\end{figure}
\vspace{-20pt}
\begin{figure}[h!]
         \centering
         \includegraphics[width=\textwidth]{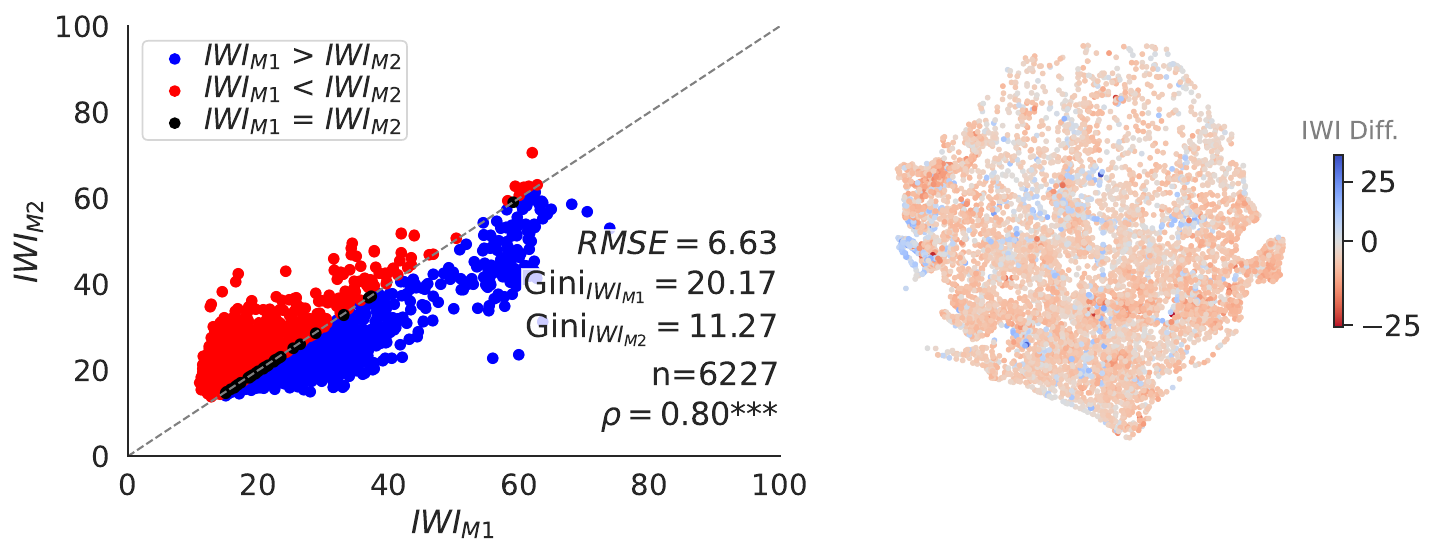}
        \vspace{-19pt}
        \caption{\textbf{IWI predictions in overlapping places in SLE ({M2}: CB).} 
        Right: Prediction differences between {M1}~and {M2}. %
        Left: Correlation between these predictions.
        Compared to {M2}, {M1}~underestimates wealth (mostly red) and predicts more inequality (higher Gini coefficient). 
        }
        \label{fig:ol_SLE}
\end{figure}
\clearpage
\newpage
\subsubsection{Liberia, LBR}
~

\vspace{-10pt}
\begin{figure}[h!]
     \centering
     \begin{subfigure}[b]{0.6\textwidth}
         \centering
         \includegraphics[width=\textwidth]{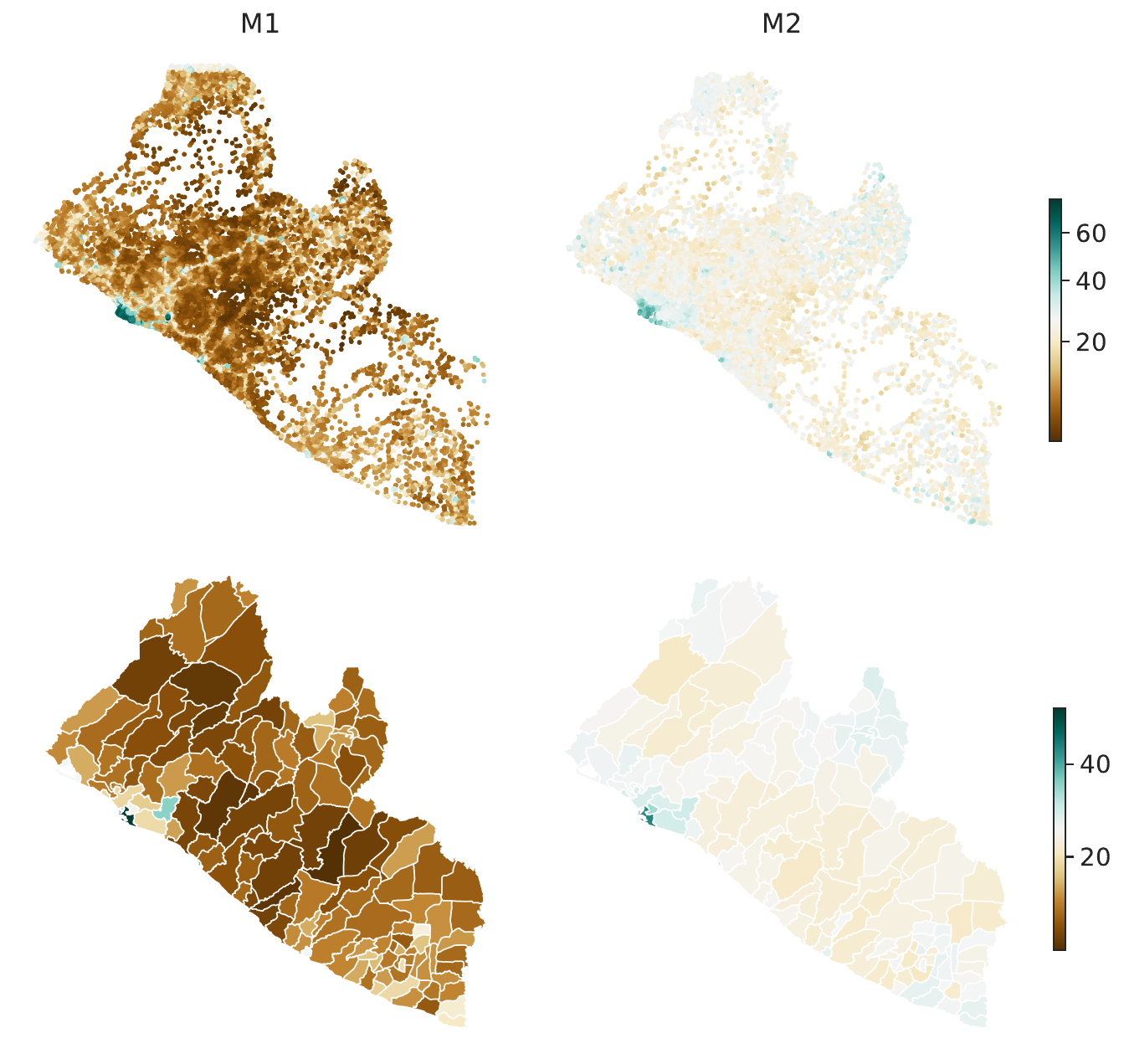}
         \caption{Poverty Map}
         \label{fig:LBR_hr}
     \end{subfigure}
     \begin{subfigure}[b]{0.39\textwidth}
         \centering
         \includegraphics[width=\textwidth]{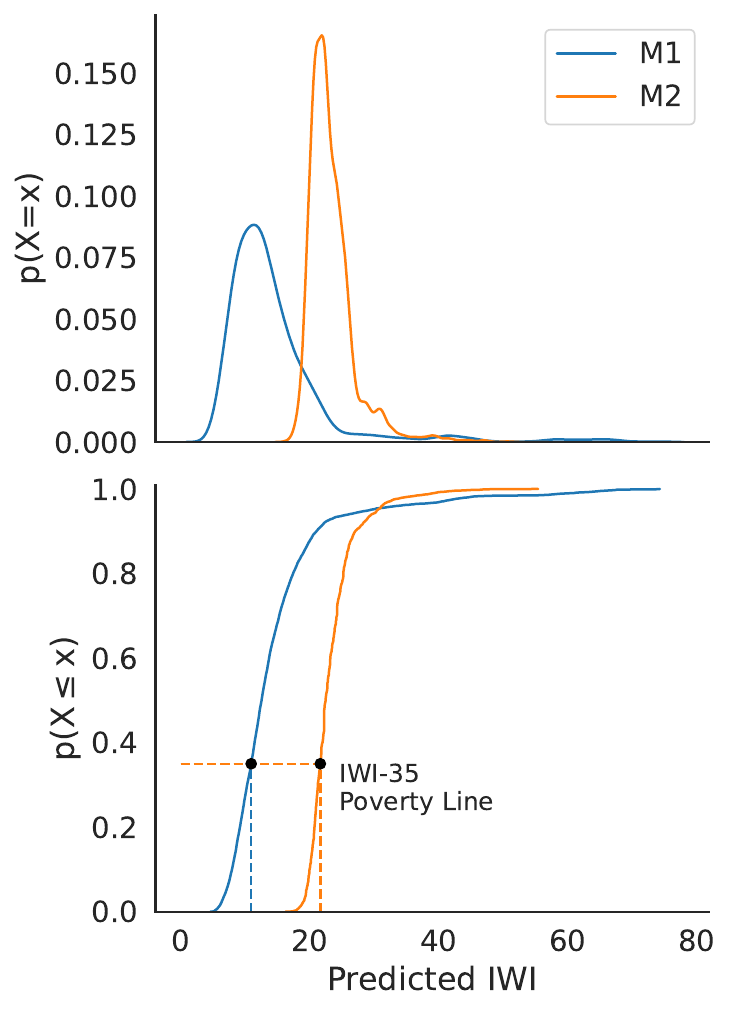}
         \caption{Wealth distribution}
         \label{fig:LBR_admin}
     \end{subfigure}
        \caption{\textbf{IWI predictions in LBR ({M2}: CB).} 
        (a) Poverty maps with colors representing the average predicted IWI scores. %
        Top: OSM populated places. %
        Bottom: District level. %
        {M1}~predicts lower wealth than {M2}.
        (b) Predicted wealth distribution. %
        IWI poverty line ($35^{th}$ percentile, corresponding to poverty headcount ratios at \$1.25 a day~\cite{smits2015international}) slightly differs between models.
        }
        \label{fig:LBR}
\end{figure}
\vspace{-20pt}
\begin{figure}[h!]
         \centering
         \includegraphics[width=\textwidth]{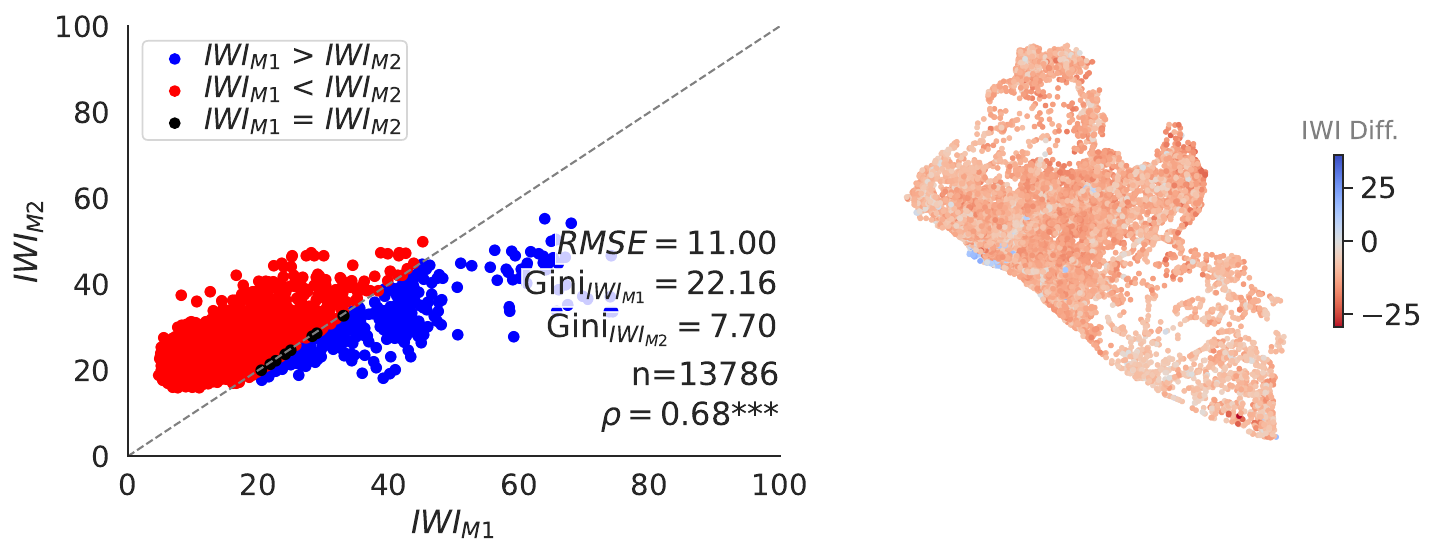}
        \vspace{-19pt}
        \caption{\textbf{IWI predictions in overlapping places in LBR ({M2}: CB).} 
        Right: Prediction differences between {M1}~and {M2}. %
        Left: Correlation between these predictions.
        Compared to {M2}, {M1}~underestimates wealth (mostly red) and predicts more inequality (higher Gini coefficient). 
        }
        \label{fig:ol_LBR}
\end{figure}
\clearpage
\newpage
\subsubsection{Uganda, UGA}
~
\vspace{-10pt}
\begin{figure}[h!]
     \centering
     \begin{subfigure}[b]{0.6\textwidth}
         \centering
         \includegraphics[width=\textwidth]{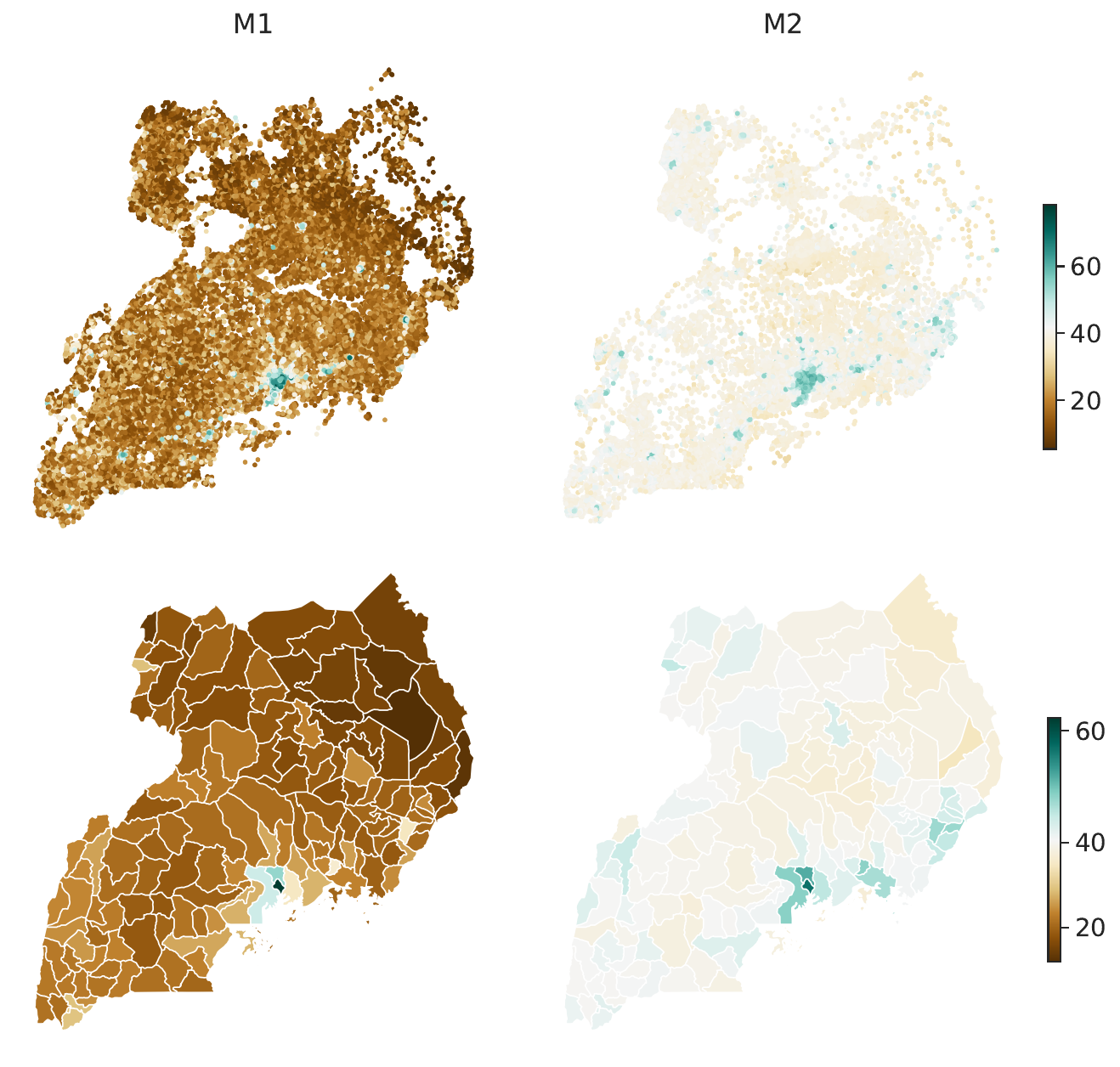}
         \caption{Poverty Map}
         \label{fig:UGA_hr}
     \end{subfigure}
     \begin{subfigure}[b]{0.39\textwidth}
         \centering
         \includegraphics[width=\textwidth]{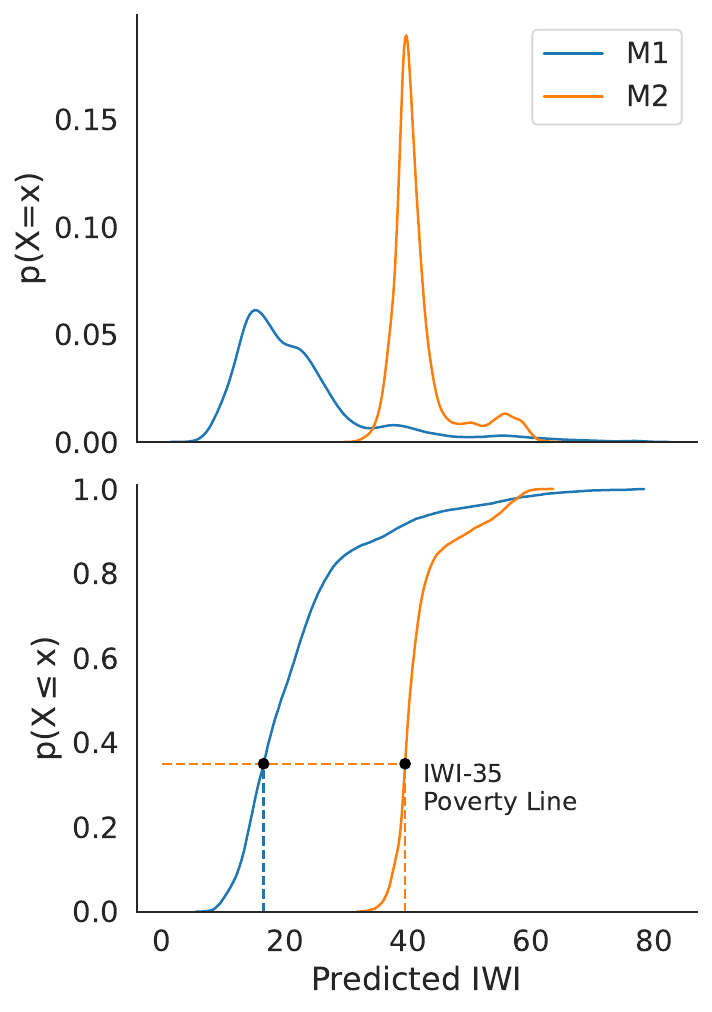}
         \caption{Wealth distribution}
         \label{fig:UGA_admin}
     \end{subfigure}
        \caption{\textbf{IWI predictions in UGA ({M2}: CB$_w$).} 
        (a) Poverty maps with colors representing the average predicted IWI scores. %
        Top: OSM populated places. %
        Bottom: District level. %
        {M1}~predicts lower wealth than {M2}.
        (b) Predicted wealth distribution. %
        IWI poverty line ($35^{th}$ percentile, corresponding to poverty headcount ratios at \$1.25 a day~\cite{smits2015international}) differs between models.
        }
        \label{fig:UGA}
\end{figure}
\vspace{-20pt}
\begin{figure}[h!]
         \centering
         \includegraphics[width=\textwidth]{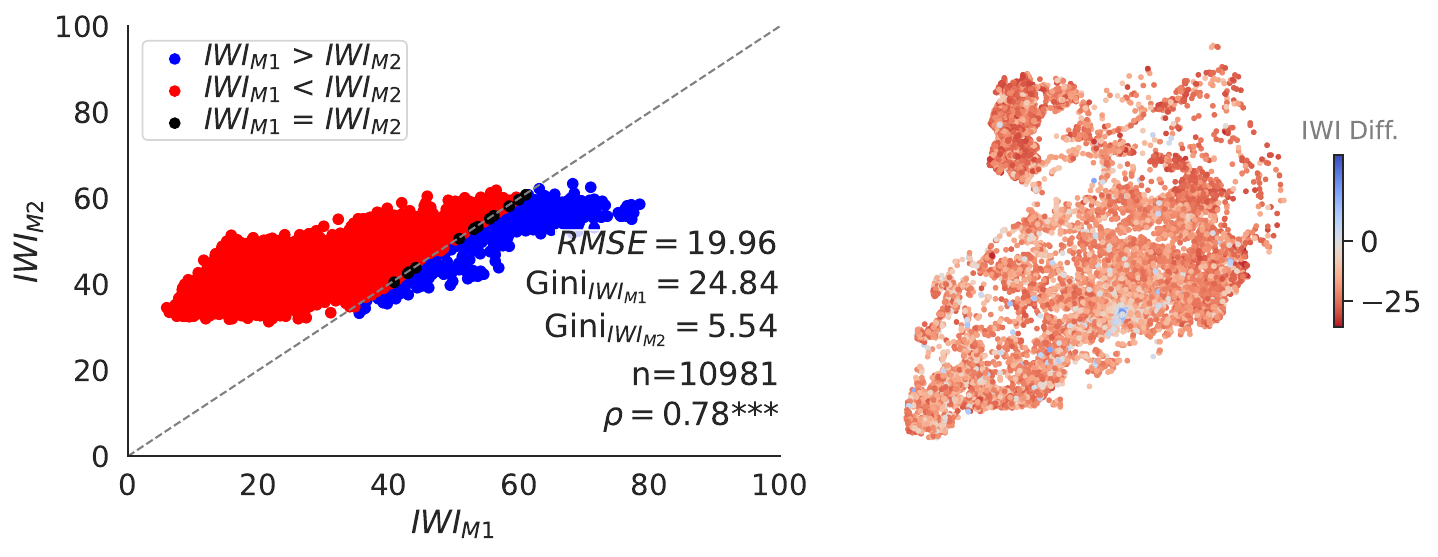}
        \vspace{-19pt}
        \caption{\textbf{IWI predictions in overlapping places in UGA ({M2}: CB$_w$).} 
        Right: Prediction differences between {M1}~and {M2}. %
        Left: Correlation between these predictions.
        Compared to {M2}, {M1}~underestimates wealth (mostly red) and predicts more inequality (higher Gini coefficient). 
        }
        \label{fig:ol_UGA}
\end{figure}
\clearpage
\newpage
\subsubsection{Rwanda, RWA}
~
\vspace{-10pt}
\begin{figure}[h!]
     \centering
     \begin{subfigure}[b]{0.6\textwidth}
         \centering
         \includegraphics[width=\textwidth]{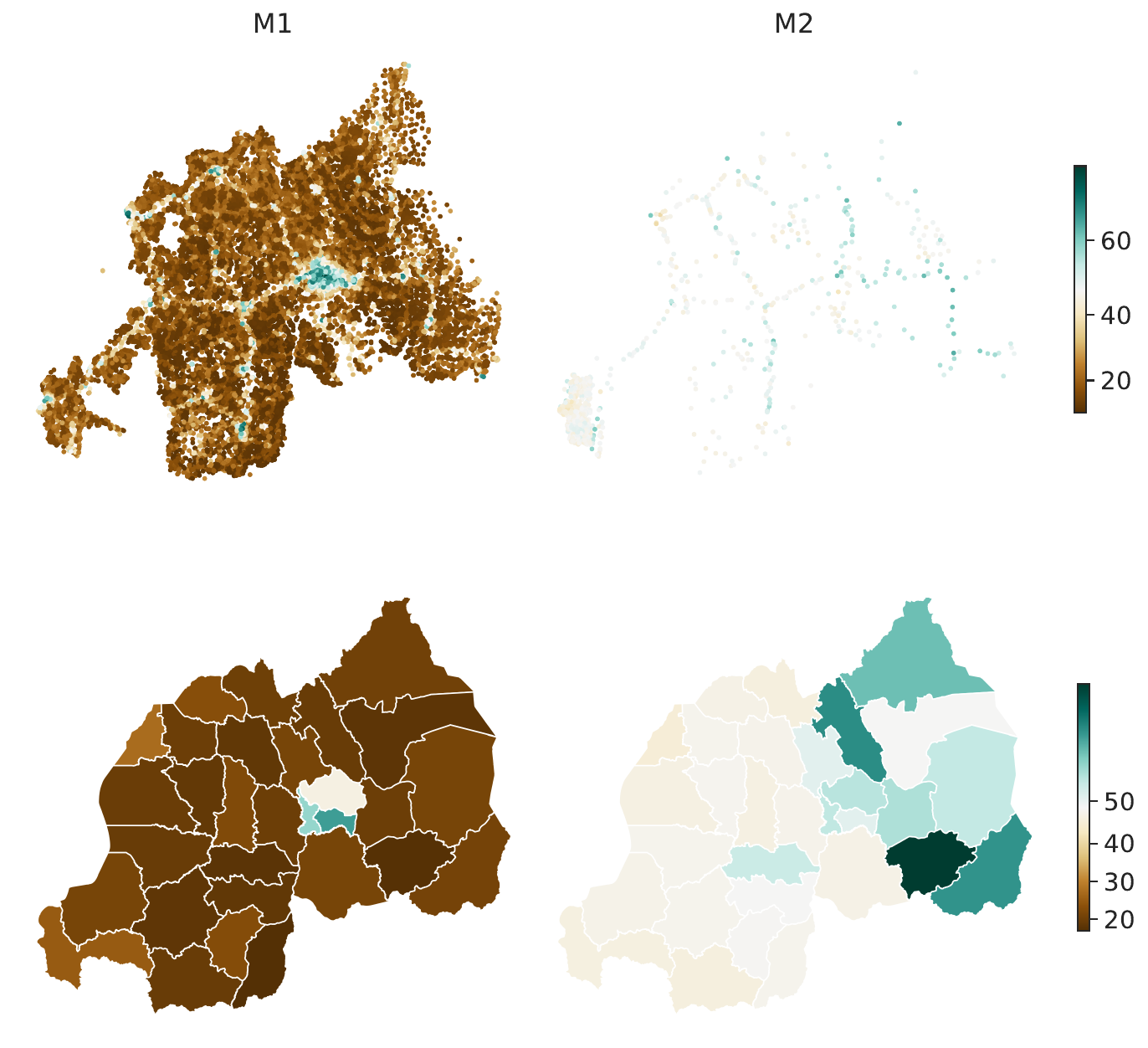}
         \caption{Poverty Map}
         \label{fig:RWA_hr}
     \end{subfigure}
     \begin{subfigure}[b]{0.39\textwidth}
         \centering
         \includegraphics[width=\textwidth]{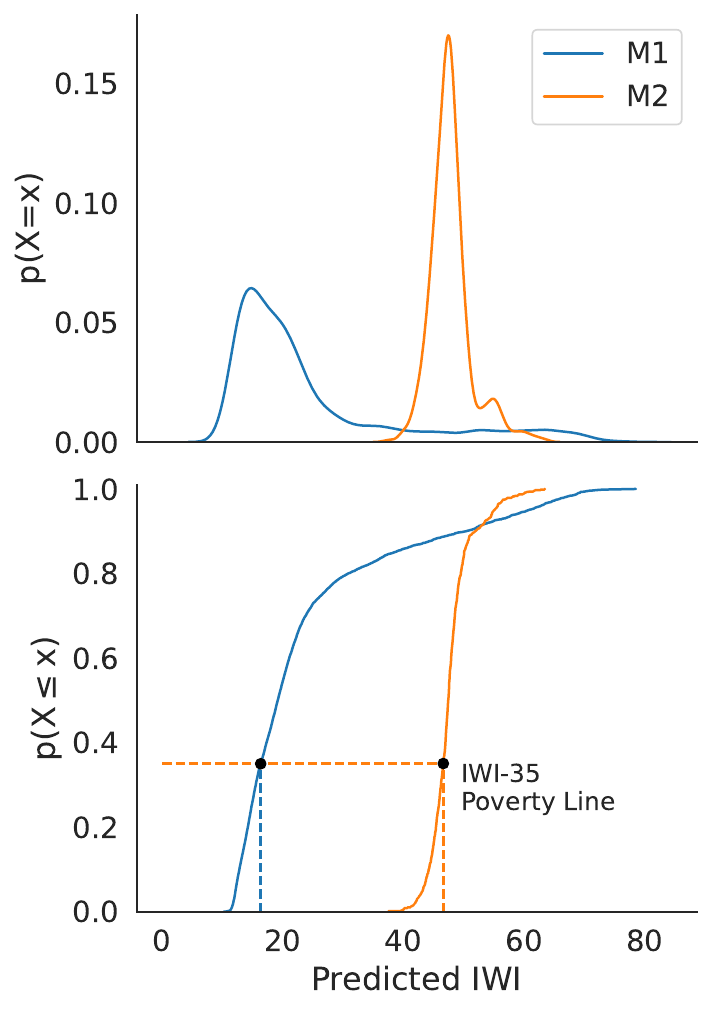}
         \caption{Wealth distribution}
         \label{fig:RWA_admin}
     \end{subfigure}
        \caption{\textbf{IWI predictions in RWA ({M2}: CB$_w$).} 
        (a) Poverty maps with colors representing the average predicted IWI scores. %
        Top: OSM populated places. %
        Bottom: District level. %
        {M1}~predicts lower wealth than {M2}.
        (b) Predicted wealth distribution. %
        IWI poverty line ($35^{th}$ percentile, corresponding to poverty headcount ratios at \$1.25 a day~\cite{smits2015international}) differs between models.
        }
        \label{fig:RWA}
\end{figure}
\vspace{-20pt}
\begin{figure}[h!]
         \centering
         \includegraphics[width=\textwidth]{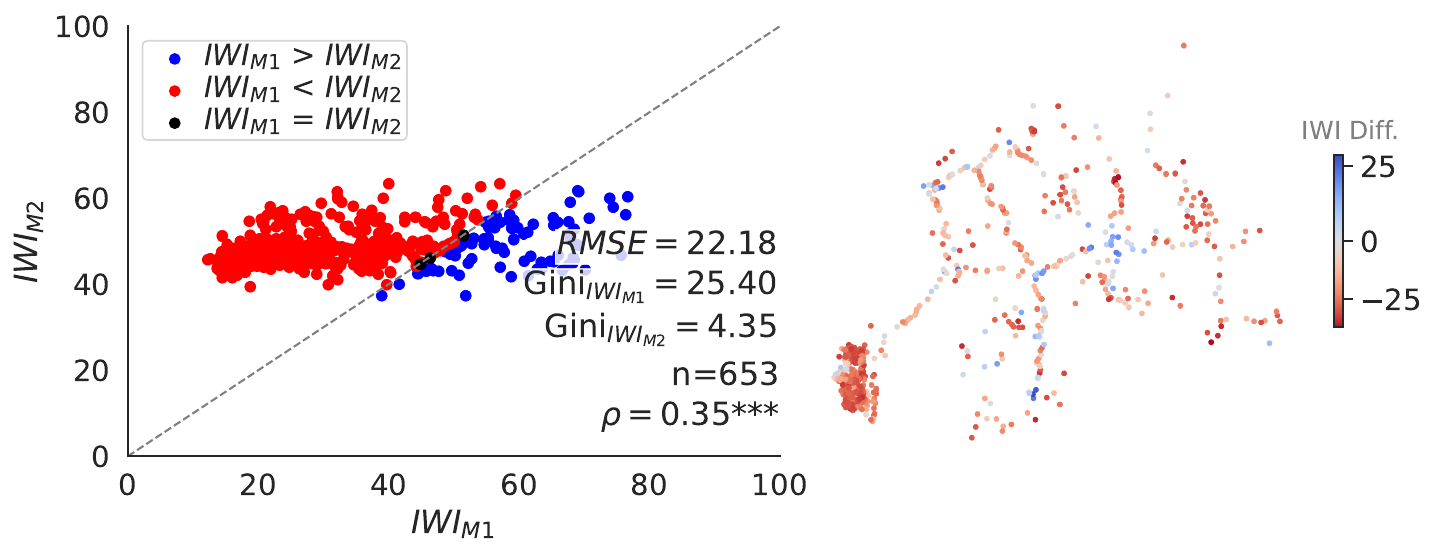}
        \vspace{-19pt}
        \caption{\textbf{IWI predictions in overlapping places in RWA ({M2}: CB$_w$).} 
        Right: Prediction differences between {M1}~and {M2}. %
        Left: Correlation between these predictions.
        Compared to {M2}, {M1}~underestimates wealth (mostly red) and predicts more inequality (higher Gini coefficient). 
        }
        \label{fig:ol_RWA}
\end{figure}
\clearpage
\newpage
\subsubsection{Gabon, GAB}
~
\vspace{-10pt}
\begin{figure}[h!]
     \centering
     \begin{subfigure}[b]{0.6\textwidth}
         \centering
         \includegraphics[width=\textwidth]{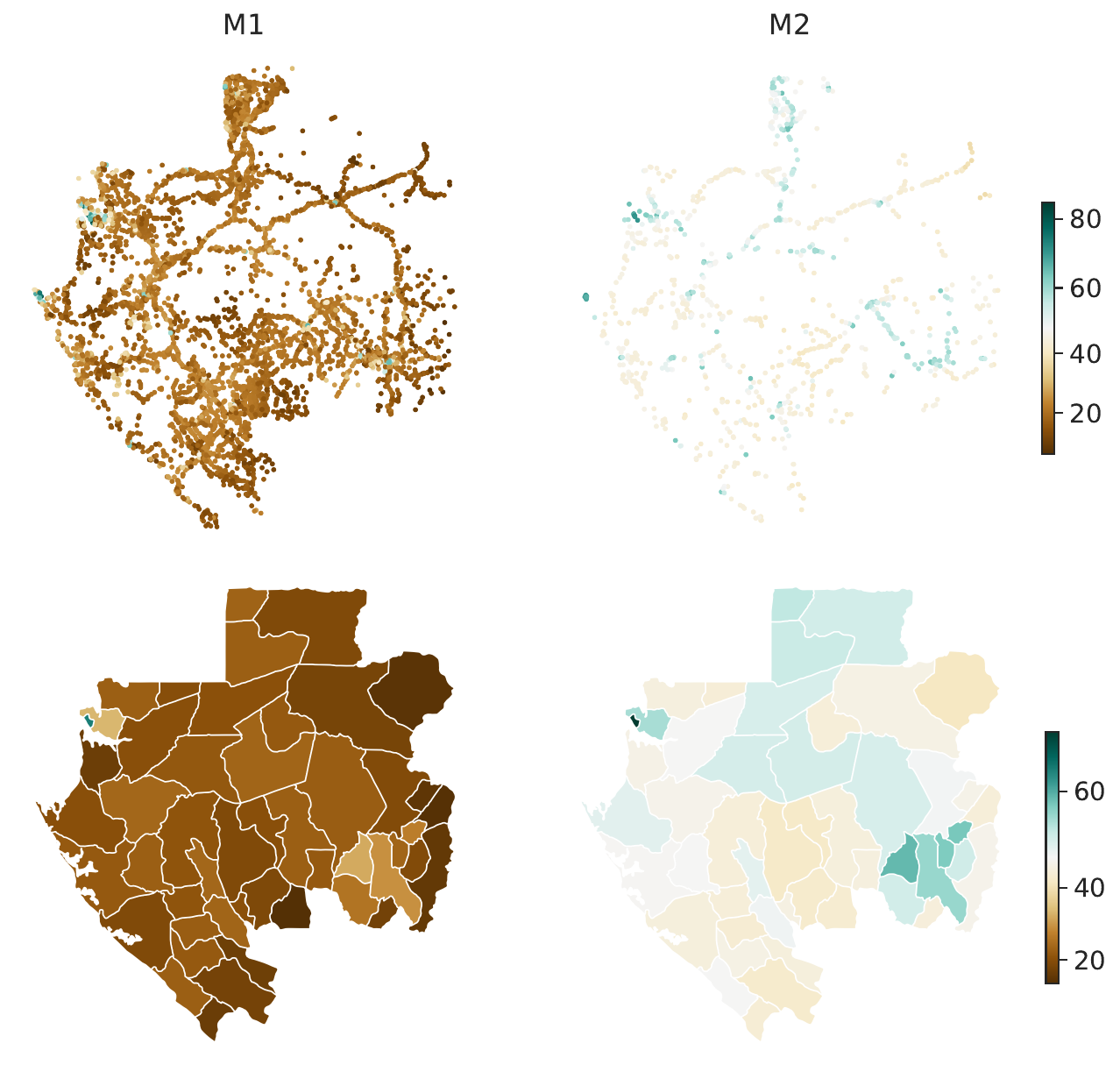}
         \caption{Poverty Map}
         \label{fig:GAB_hr}
     \end{subfigure}
     \begin{subfigure}[b]{0.39\textwidth}
         \centering
         \includegraphics[width=\textwidth]{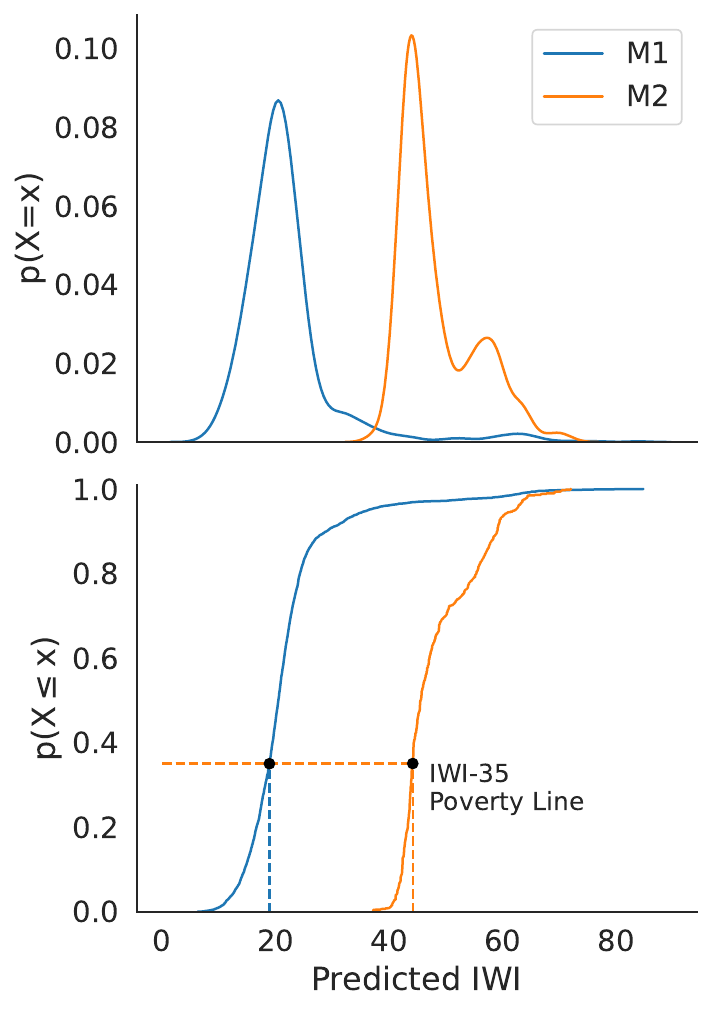}
         \caption{Wealth distribution}
         \label{fig:GAB_admin}
     \end{subfigure}
        \caption{\textbf{IWI predictions in GAB ({M2}: CNN$_a$+CB).} 
        (a) Poverty maps with colors representing the average predicted IWI scores. %
        Top: OSM populated places. %
        Bottom: District level. %
        {M1}~predicts lower wealth than {M2}.
        (b) Predicted wealth distribution. %
        IWI poverty line ($35^{th}$ percentile, corresponding to poverty headcount ratios at \$1.25 a day~\cite{smits2015international}) differs between models.
        }
        \label{fig:GAB}
\end{figure}
\vspace{-20pt}
\begin{figure}[h!]
         \centering
         \includegraphics[width=\textwidth]{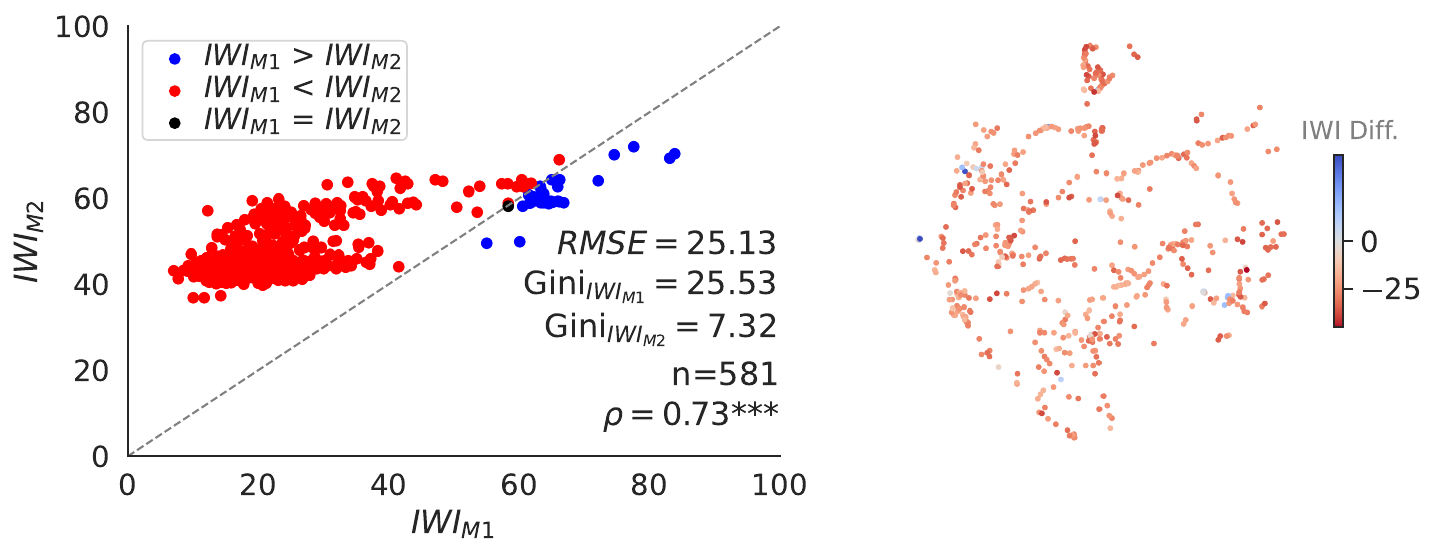}
        \vspace{-19pt}
        \caption{\textbf{IWI predictions in overlapping places in GAB ({M2}: CNN$_a$+CB).} 
        Right: Prediction differences between {M1}~and {M2}. %
        Left: Correlation between these predictions.
        Compared to {M2}, {M1}~underestimates wealth (mostly red) and predicts more inequality (higher Gini coefficient). 
        }
        \label{fig:ol_GAB}
\end{figure}\subsection{Performance on test sets}
\label{app:performance}
In~\Cref{tbl:performance}, we present the reported evaluation scores (R$^2$) on the test set for each model across different countries, provided solely for background context. It is important to note that these scores are not directly comparable due to several factors.
First, each model was trained on data from different years and places, leading to variations in the socioeconomic conditions captured.
Even in cases like UGA and ZAF, where {M1}~and {M2}~used data from the same year, their training, validation, and test sets were not identical.
Second, while {M1}~and {M2}~predict IWI (International Wealth Index) scores, M3 predicts RWI (Relative Wealth Index) scores, further complicating a direct comparison. 
Therefore, the R$^2$ values in~\Cref{tbl:performance} should be seen as individual performance metrics rather than for cross-model comparison.

\begin{table}[h]
\centering
\caption{\textbf{Reported validation R$^2$ scores for each approach.} 
{M1}~presents validation results for single-country training in Table 1~\cite{lee2022high}. 
{M3}~reports validation results on census data, as shown in Fig. S4 of the supplementary information~\cite{chi2022microestimates}. 
We retrain the {M2}~models for each country using a subset of the original features reported in~\cite{espin2023interpreting}, and report the performance (R$^2$) of the best model, indicated by the lowest root-mean-square error, RMSE (see details in~\Cref{sec:models}).}
\label{tbl:performance}
\begin{tabular}{@{}p{3.cm}%
>{\raggedleft\arraybackslash}p{1.3cm}%
>{\raggedleft\arraybackslash}p{1.3cm}%
>{\raggedleft\arraybackslash}p{1.3cm}@{}}
\toprule
Country (Code) / R$^2$  & {M1} & {M2} & {M3}\\ \midrule
Sierra Leone (SLE)  & 89.82 & 80.7 & 83.0 \\
Liberia (LBR)  & 89.3 & 76.3 & - \\
Uganda (UGA)  & 89.02 & 78.5 & - \\
Rwanda (RWA)  & 87.41 & 67.5 & 82.0 \\
South Africa (ZAF)  & 73.03 & 49.5 & -  \\
Gabon (GAB) & - & 81.9 & - \\ \bottomrule
\end{tabular}
\end{table}\begin{figure}[t]
    \centering
    \includegraphics[width=1.\linewidth]{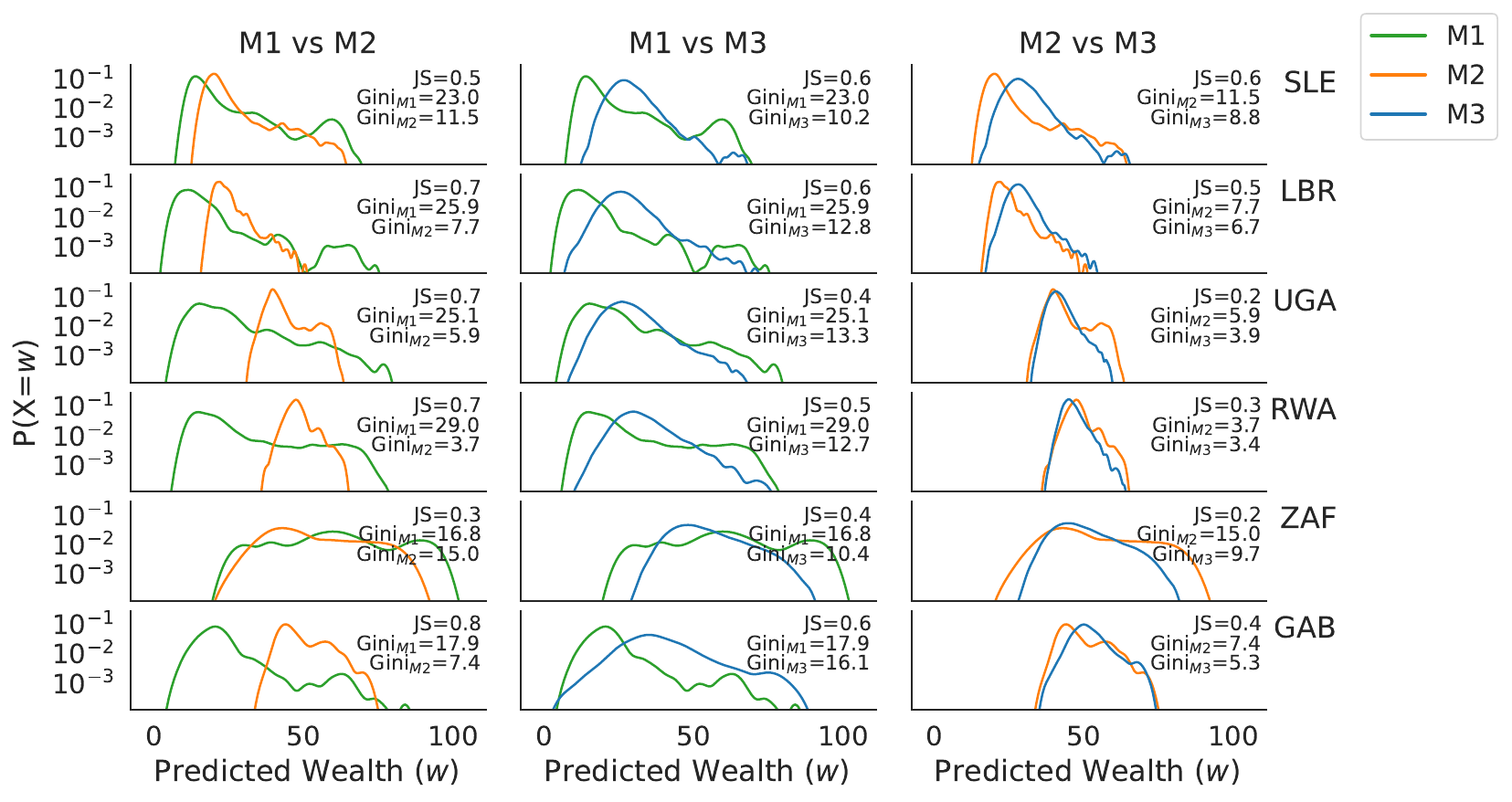}
    \caption{\textbf{Predicted wealth distributions}. Every column compares the predicted wealth distributions between two models.
    {M1}~and {M2}~predict IWI scores, while {M3}~predicts RWI scores. 
    For comparisons with {M3}, we apply a linear transformation from RWI to IWI using~\Cref{eq:rescale}, with {M1}~and {M2}~as baselines in the middle and rightmost columns, respectively.
    {M3}'s distribution (blue) aligns more with {M2}~(orange), as indicated by a lower Jensen-Shannon (JS) distance, suggesting similar skewness levels (Gini coefficient) between them.
    }
    \label{fig:comparison-all}
\end{figure}

\subsection{Comparison against {M3}, Chi et al. (2022)}
\label{app:chi}
The Relative Wealth Index (RWI) provides relative wealth scores that allow for comparisons within a given country, offering insights into spatial wealth disparities. However, a key limitation of the RWI is its lack of comparability across countries and over time, as it reflects relative wealth only within the specific context of the data used either from survey data or to train ML models. Consequently, RWI scores are not directly comparable across different geographic regions or time periods due to variations in the underlying data and economic conditions.
To address this limitation, some studies have proposed transformations of the RWI to create a more standardized and comparable index across countries and years. One such proposal is the Comparable Wealth Index (CWI)~\cite{rutstein2014making}, which aims to standardize wealth measurement globally. The CWI adjusts RWI scores to account for cross-national differences, allowing for more meaningful comparisons across different socioeconomic contexts. For a detailed comparison of wealth scores derived from survey data, see~\cite{mayfour2022assessing}.

Although conversion methods between wealth indicators have been proposed, they commonly rely on reference measures that are not easily available. For this reason here we decided to follow a simple approach to fairly compare different methods. First, we make the assumption that {M3}~(which uses RWI) correlates strongly with {M1}~(which uses IWI), as evidenced by the very high correlation between RWI and IWI on survey data (Pearson's correlation: $0.93-0.99$~\cite{smits2015international}). Under this assumption, we apply a linear transformation of RWI to IWI using~\Cref{eq:rescale}, enabling us to make relative comparisons between {M1}~and {M3}. Similarly, we apply the transformation when comparing {M2}~to {M3}~using {M2}~as a baseline. While this transformation is not ideal and may introduce some inaccuracies, it serves as a useful tool for comparing the shape of the predicted distributions. These results should therefore be interpreted with caution.

\begin{equation}
    \hat{iwi} = \frac{(rwi - rwi_{min})*(iwi_{max}-iwi_{min})}{(rwi_{max}-rwi_{min})}+iwi_{min}
\label{eq:rescale}
\end{equation}

\Cref{fig:comparison-all} shows the comparison of predicted wealth distributions across models. 
We observe that the shape of {M3}'s distributions and its wealth inequality more closely align with {M2}, as indicated by a smaller Jensen-Shannon distance~\cite{lin1991divergence} and similar Gini coefficient values compared to {M1}, suggesting that both models ({M3}~and {M2}) produce similar distributions of wealth.
One possible explanation for this behavior is the similarity in the models' architectures and the features they use, see~\Cref{tbl:ground_truth}. 
Note that to preserve the integrity of the data provided by the three models, we chose to retain outliers in the distributions. This decision results in apparent mismatches between the minimum and maximum values across distributions in~\Cref{fig:comparison-all}. Removing these outliers would lead to a more favorable but unfair comparison of the altered distributions.

\end{document}